\documentclass[10pt,draftclsnofoot,onecolumn]{IEEEtran}
\usepackage[utf8]{inputenc}
\usepackage{amsmath}
\usepackage{amssymb}
\usepackage{upgreek}
\usepackage{tipa}
\usepackage{dsfont}
\usepackage{enumerate}
\usepackage{subfig}
\usepackage{float}
\usepackage{multicol}
\usepackage[usenames,dvipsnames]{pstricks}
\usepackage{pst-grad}
\usepackage{pst-plot}
\usepackage[noadjust]{cite}
\usepackage{graphicx}
\usepackage{epstopdf}
\usepackage{psfrag}
\usepackage{booktabs}
\begin{document}

\title{Stability, convergence, and limit cycles in some human physiological processes}

\author{\IEEEauthorblockN{Sreelakshmi Manjunath, Gopal Krishna Kamath and Gaurav Raina} \\
\IEEEauthorblockA{Department of Electrical Engineering, Indian Institute of Technology Madras, Chennai 600 036, India\\
Email: $\lbrace \text{sreelakshmi, ee12d033, gaurav} \rbrace$@ee.iitm.ac.in}
\thanks{This is an extension of our preliminary work that appeared in Proceedings of the 26th IEEE Chinese Control and Decision Conference (CCDC), pp. 2076-2082,
2014. DOI: 10.1109/CCDC.2014.6852509}
}

%




\maketitle

\begin{abstract}
Mathematical models for physiological processes aid qualitative understanding of the impact of various parameters on the underlying process. 
We analyse two such models for human physiological processes: the Mackey-Glass and the Lasota equations, which model the change in the concentration of blood cells in the human body. We first study the local stability of these models, and derive bounds on various model parameters and the feedback delay for the concentration to equilibrate. We then deduce conditions for non-oscillatory convergence of the solutions, which could ensure that the blood cell concentration does not oscillate. Further, we define the convergence characteristics of the solutions which govern the rate at which the concentration equilibrates when the system is stable. Owing to the possibility that physiological parameters can seldom be estimated precisely, we also derive bounds for robust stability\textemdash which enable one to ensure that the blood cell concentration equilibrates despite parametric uncertainty. We also highlight that when the necessary and sufficient condition for local 
stability is violated, the system transits into instability via a Hopf bifurcation, leading to limit cycles in the blood cell concentration. We then outline a framework to characterise the type of the Hopf bifurcation and determine the asymptotic orbital stability of limit cycles. The analysis is complemented with numerical examples, stability charts and bifurcation diagrams. The insights into the dynamical properties of the mathematical models may serve to guide the study of dynamical diseases.

\end{abstract}

\section{Introduction}
\label{sec:introduction}
Mathematical models for human physiological processes can help gain insight into underlying phenomena. Environmental factors, or structural changes in the phenomena can be captured by model parameters, and the impact of variation in these parameters on the physiological process can be explored. Analysis of such models enables one to understand the change in physiological rhythms, and predict the onset of pathological behaviour in the form of disturbances in these rhythms~\cite{Macky_77, Hennani_15}. In addition to this, studying these models has lead to various physiological and mathematical theories~\cite{Keener_98, Ottesen_04}. Therefore, analysing these models has been an area of interest for biologists and mathematicians alike. For some recent investigations; see~\cite{Bate_13,Diekmann_10, Foley_09,Karst_15,Langlois_17,Serna_13,Yanchuk_17,Scharf_17,Shu_17}.

It is well known that initial symptoms of physiological diseases manifest as irregularity in periodic rhythms; such as a lack of or a change in the periodicity of otherwise periodic phenomena, or the onset of oscillations in non-oscillatory processes~\cite{Murray_07,Keener_98}. For example, the onset of oscillations in the count of white blood cells could indicate symptoms of leukaemia~\cite{Foley_09}. Such physiological diseases, that can be characterised by a qualitative change in dynamics, are termed as \emph{dynamical diseases}~\cite{Belair_95,Mackey_87}. Analysing mathematical models to predict this qualitative change, and devising methods for avoiding such changes could eventually guide therapy for dynamical diseases~\cite{Macky79a}.

It has been recognised that physiological processes can be modelled by dynamical  systems \cite{Macky_77,Lasota_77,Glass_Beuter_88,Waz_76}. Such models typically incorporate finite time delays in the feedback loops, as they provide a more accurate description of the systems they represent. To understand the theoretical properties of the models, a control- and a bifurcation-theoretic analysis should be pursued. A control-theoretic investigation can reveal trade-offs in system design, and the choice of model parameters, that help to maintain stability. A bifurcation-theoretic analysis, on the other hand, can alert us about the onset of sophisticated dynamical behaviour as model parameters may vary. In fact, researchers often associate bifurcations in mathematical models of physiological process with abnormal dynamics or diseases~\cite{Glass_79,Milton_13,Beuter_95,Haurie_98}.
Thus, a combination of control and bifurcation theory can provide a systematic understanding of the system dynamics. In this paper, we investigate some stability and local bifurcation properties of two models from physiology.  

In the wide class of physiological models, we choose to study the Mackey-Glass~\cite{Macky_77} and the Lasota~\cite{Lasota_77} equations.
These equations are non-linear delay differential equations that can display a broad range of dynamical behaviour. The Mackey-Glass equation has been extensively studied, but still continues to generate theoretical interest; for example, see~\cite{Baker_14} and references therein. The Lasota equation, in contrast, has received much less attention from the theoretical community. However, there is recent and renewed interest in the Lasota equation. Stability, for the Mackey-Glass and Lasota equations, in the sense of Hyers and Ulam has been recently reported~\cite{Qarawani_13}. Also, by employing an ergodic theory approach to chaos, Lasota's conjecture about non-trivial ergodic properties of the model has been explored in~\cite{Mitkowski_12}. Given the continued interest in the Mackey-Glass model, and renewed interest in the Lasota equation, we conduct a systematic local stability and a local Hopf bifurcation analysis of both these models. 
Typically, it is highlighted that time delayed equations may lose stability as the feedback delay increases beyond a threshold value~\cite{Hassard_81,Glass_Beuter_88}. 
Furthermore, variation in any of the model parameters may also destabilise the system. This implies that, either the time delay or any of the model parameters can act as the bifurcation parameter. However, as either of these parameters varies, the system equilibrium also shifts, thus making it cumbersome to study the change in system dynamics. 
To that end, in this paper, we employ a non-dimensional bifurcation parameter to conduct the bifurcation analysis. To the best of our knowledge, a detailed local Hopf bifurcation analysis of the Lasota equation has not been conducted before.                

For both physiological models, we first conduct a local stability analysis. We first outline a sufficient condition and the necessary and sufficient condition for local stability. Sufficient conditions for stability can possibly aid in the design of system parameters to ensure stable operation. We then show that violating the necessary and sufficient condition for local stability results in a Hopf bifurcation~\cite{Guckenheimer_83,Kuznetsov_13}.
Such conditions establish bounds on system parameters and the feedback delay to maintain stability. We also outline the conditions on system parameters and the feedback delay to ensure non-oscillatory convergence of the solutions to the system equilibrium. Further, since the model parameters in these equations need to be estimated from data, there is an uncertainty (called `parametric uncertainty') which is naturally introduced in their estimated values. Hence, we also outline bounds on the feedback delay that would ensure local stability despite the presence of parametric uncertainties.

As variation in the feedback delay or any of the system parameters can destabilise the system, we conduct the bifurcation analysis via a suitably motivated non-dimensional parameter. The Hopf condition informs us that we could expect to see a limit cycle branching from the fixed point. Using Poincar\'{e} normal forms and the center manifold theory \cite{Hassard_81}, we also provide the requisite theoretical analysis to determine the type of the Hopf bifurcation. The analysis, in this paper, is complemented with numerical examples, stability charts and bifurcation diagrams.           

The rest of this paper is organised as follows. In Section~\ref{sec:models}, we outline the Mackey-Glass and Lasota equations. In Section~\ref{sec:local_stability}, we outline a sufficient condition, and the necessary and sufficient condition, for local stability and derive the condition for non-oscillatory convergence. In Section~\ref{sec:roc}, we study the convergence properties of the two models. Section~\ref{sec:robust_stability} comprises the robust stability analysis for both the models. We then conduct a local Hopf bifurcation analysis in Section~\ref{sec:Hopf_bifurcation}. Finally, in Section~\ref{sec:contributions}, we summarise our contributions. Appendix A contains the local stability analysis of a general first order non-linear time-delayed system, and in Appendix B we outline the local Hopf bifurcation analysis.     

\section{Models}   
\label{sec:models}
Models for physiological processes can be used to predict periodic and aperiodic dynamics that may be found in human diseases. Haematopoiesis, the process of production of blood cells, 
has been modelled via a delay differential equation \cite{Macky_77}. It has been shown that such models can reproduce the qualitative features of both normal and pathological behaviour. In a similar spirit, erythropoiesis, the process of formation of red blood cells in the bone marrow, has also been modelled via a dynamical systems approach \cite{Lasota_77}. In order to understand both normal and pathological behaviour, a control and a bifurcation theoretic approach is desirable. In this section, we describe the Mackey-Glass \cite{Macky_77} and Lasota \cite{Lasota_77} equations, which 
are the models we study in this paper.    

\subsection{Mackey-Glass equation}
The Mackey-Glass equation \cite{Macky_77} considers a population of mature circulating cells, and models the process of haematopoiesis by the following dynamical system 
\begin{align}
 \dot{x}(t) =\beta \frac{\tilde{\theta}^n x(t-\tau)}{\tilde{\theta}^n + x^n(t-\tau)} - \gamma x(t),\label{eq:m-g}
\end{align}
where $x(t)$ (cells/kg) represents the concentration of mature blood cells in circulation; and throughout this paper, we use the notation, $\dot{x}(t)= dx/dt$. The constants $\beta$ (day$^{-1}$) and $\tilde{\theta}$ (cells/kg) capture the dependence of the production of blood cells on the number of mature cells in the blood stream. The parameter $\gamma$ (day$^{-1}$) represents the rate at which cells are lost from circulation in proportional to the concentration of the circulating cells.
The parameter $n$ is intended to capture the non-linearity in the haematopoietic process. There is a finite time delay, $\tau$, between the initiation of production of blood cells in the bone marrow and the release of mature cells into the blood stream. The dependence of $\dot{x}(t)$ on $x(t-\tau)$ is intended to capture the effects of poietin feedback control from the circulating population of cells~\cite{Glass_79}. The parameters $\beta,\tilde{\theta}, n, \gamma$, and the time delay $\tau$ are strictly positive.

In a well studied variant of the Mackey-Glass equation it is assumed that $\tilde{\theta}=1$ \cite{Qarawani_13}. In this case, equation \eqref{eq:m-g} becomes
\begin{align}
 \dot{x}(t) =\beta \frac{x(t-\tau)}{1 + x^n(t-\tau)} - \gamma x(t).\label{eq:mod-m-g}
\end{align}
We refer to \eqref{eq:mod-m-g} as the Mackey-Glass equation.
\subsection{Lasota equation}
The Lasota equation \cite{Lasota_77} models the dynamics of the formation of red blood cells in the bone marrow.
The corresponding dynamical system is
\begin{align}
\dot{x}(t) = \beta x^n(t-\tau) e^{-x(t-\tau)}-\gamma x(t),\label{eq:lasota}
\end{align}
where $x(t)$ represents the number of erythrocytes, \emph{i.e.,} the red blood cells in blood circulation. The constants $\beta,n,\gamma,\tau > 0$. The parameter $\beta$ captures the demand for oxygen. The process of erythropoiesis is inherently non-linear in nature, and the parameter $n$ is intended to capture the proportional part of this non-linearity. The rate of destruction of the blood cells is represented by $\gamma$. The time delay, $\tau$, which greatly influences the system dynamics, represents the time required for the erythrocytes to attain maturity.

  
\section{Local Stability}
\label{sec:local_stability}
Observe that both the Mackey-Glass and Lasota equations are non-linear, time delayed models of the form
\begin{align}
 \dot{x}(t) =\,\,\beta F\big(x(t-\tau)\big) -\gamma x(t),\label{eq:general}
\end{align}
where $F(\cdot)$ is a non-linear function of the blood-cell concentration $\tau$ units of time prior to the current time instant.
For the analysis of such non-linear models, we first usually conduct a local stability analysis. 
Let the non-trivial equilibrium of equation \eqref{eq:general} be $x^\ast$, then the equilibrium satisfies
\begin{align}
\beta F(x^{\ast}) = \gamma x^{\ast}.\label{eq:gen-eq-cond}
\end{align}
Let $u(t)=x(t)-x^\ast$ denote a perturbation about the equilibrium $x^{\ast}$. 
Then the linearised system, associated with equation \eqref{eq:general}, is
\begin{align}
 \dot{u}(t) = -a u(t) - bu(t-\tau),\label{eq:general-lin}
\end{align}
where 
\begin{align}
 &a = \gamma > 0,
 & b = -\beta F'(x^\ast)>0.\label{eq:ab}
\end{align}
In this paper, we focus on the case where $a\geq0,b>0$ and $b>a$. Looking for exponential solutions of~\eqref{eq:general-lin} we get the characteristic equation
\begin{align}
 \lambda + a + be^{-\lambda\tau} = 0.\label{eq:general-char}
\end{align}
From the stability analysis of equation \eqref{eq:general-lin}, outlined in Appendix A, a sufficient condition for stability is $b\tau<\pi/2$. Using this condition, a \emph{sufficient condition for local stability} of equation \eqref{eq:general} is 
\begin{align}
  -\,F'(x^\ast)\beta\tau< \frac{\pi}{2}.\label{eq:cond-suff}
\end{align}
The equilibrium does not depend on the time delay, but the delay does impact stability. 
Denote the value of time delay $\tau$ for which the above inequality can be replaced with an equality as $\tau_{suff}$. Now, we may say that $\tau < \tau_{suff}$ is a sufficient condition for local stability of equation~\eqref{eq:general}.

 The necessary and sufficient condition for local stability of equation \eqref{eq:general}, using the results in Appendix A, is
\begin{align}
 \tau \sqrt{\big(\beta F'(x^\ast)\big)^2 - \gamma^2} < \cos^{-1}\bigg(\frac{\gamma}{\beta F'(x^\ast)}\bigg).\label{eq:cond-necc}
\end{align}
and the Hopf bifurcation occurs at
\begin{align}
  \tau \sqrt{\big(\beta F'(x^\ast)\big)^2 - \gamma^2} = \cos^{-1}\bigg(\frac{\gamma}{\beta F'(x^\ast)}\bigg),\label{eq:cond-hopf}
\end{align}
resulting in oscillation having a period of 
\begin{align*}
 2\pi\bigg/\sqrt{\big(\beta F'(x^\ast)\big)^2 - \gamma^2}.
\end{align*}
In this paper, we denote the critical time delay that satisfies the Hopf condition as $\tau_c$.

\begin{figure}
\centering
  \psfrag{b}[b]{\hspace{-15.5mm}$\scriptsize-\beta F'(x^\ast)$}
  \psfrag{a}{\hspace{-12.5mm}Destruction rate $\gamma$}
  \psfrag{p}{$\pi/2$}
  \psfrag{nnnnnnnnnnnnnnnnnn}{\begin{scriptsize}Necessary and Sufficient\end{scriptsize}}
  \psfrag{sssssssssssss}{\begin{scriptsize}Sufficient\end{scriptsize}}
  \psfrag{RRRRRRRRRRR}{\hspace{-2mm}\begin{scriptsize}Stable region\end{scriptsize}}
  \psfrag{0.0}[b][b]{\scriptsize$0$}
   \psfrag{1.0}[b][b]{\scriptsize$1$}
     \psfrag{2.0}[b][b]{\scriptsize$2$}
   \psfrag{1.5}[b][b]{\scriptsize$1.5$}
     \psfrag{0.5}[b][b]{\scriptsize$0.5$}
   \psfrag{1}[b][b]{\scriptsize$1$}
     \psfrag{2.5}[b][b]{\scriptsize$2.5$}
     \psfrag{3.0}[b][b]{\scriptsize$3$}
\hspace{17mm}  \includegraphics[width=1.6in,height=2.4in,angle=270]{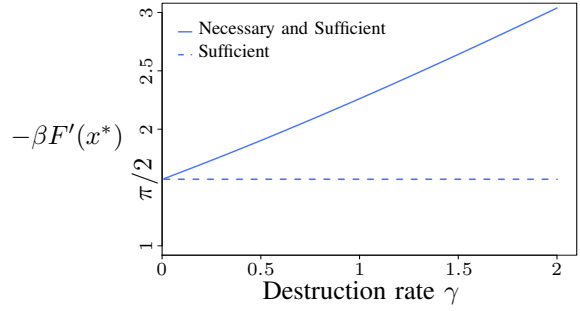}
  \caption{Local stability chart associated with the non-linear equation \eqref{eq:general}. The \emph{sufficient condition} for local stability is stated in \eqref{eq:cond-suff}, and the \emph{necessary and sufficient condition} is outlined in \eqref{eq:cond-necc}. In the above chart, $\tau=1$. The region below the lines is stable.}\label{fig:stab_chart}
\end{figure}

We have derived bounds, on the time delay and the system parameters, for local stability. 
We now seek bounds on the region, within the stable region, in which the system exhibits non-oscillatory convergence. In order to do so, we closely follow the analysis outlined in~\cite{Ghosh_arx_16}.

In the region of non-oscillatory convergence, the system parameters and the time delay are such that the characteristic equation has negative real roots. This leads to non-oscillatory convergence of the solutions to the equilibrium. For the solution of the linearised system~\eqref{eq:general-lin} to be non-oscillatory, we require the curve $f(\lambda) = \lambda + a + be^{-\lambda\tau}$ to touch the real axis. If $\sigma$ is the root of the characteristic equation at this point, we have 
\begin{align}
 f(\sigma)=&\,\sigma + a + be^{-\sigma\tau} = 0,\label{eq:f}\\
 f'(\sigma)=&\,1-b\tau e^{-\sigma\tau} = 0.\label{eq:fdash}
\end{align}
From \eqref{eq:fdash}
\begin{align}
 b\tau e^{-\sigma\tau} = 1 \hspace{3mm}\text{and}\hspace{3mm} \sigma =  \frac{\ln(b\tau)}{\tau}.\label{eq:conds}
\end{align}
Substituting \eqref{eq:conds} in \eqref{eq:f}, we have 
\begin{align*}
 \ln(b\tau) + a\tau + 1 = 0,
\end{align*}
which, on simplifying, yields
\begin{align*}
 b\tau e^{a\tau} = \frac{1}{e}.
\end{align*}
Therefore, for non-oscillatory convergence, we require
\begin{align*}
 b\tau e^{a\tau} < \frac{1}{e}.
\end{align*}
Substituting $a$ and $b$ from~\eqref{eq:ab}, we get 
\begin{align}
 -\beta F'(x^\ast)\tau e^{\gamma\tau} < \frac{1}{e}.\label{eq:non-oscill-gen}
\end{align}
Let $\tau_{noc}$ be the value of the time delay at which 
\begin{align*}
 -\beta F'(x^\ast)\tau_{noc} e^{\gamma\tau} = \frac{1}{e},
\end{align*}
then the necessary and sufficient condition for non-oscillatory convergence of equation~\eqref{eq:general} can be written as $\tau < \tau_{noc}$.

The local stability chart associated with the non-linear equation \eqref{eq:general} is shown in Figure \ref{fig:stab_chart}. 
The sufficient condition for stability is in fact not very conservative. Further, the generality of this condition allows flexibility in the choice of functional forms $F$ to ensure stability. To that end, the sufficient condition may be useful in design considerations.

As noted above, the Mackey-Glass equation and the Lasota equation, basically differ in the non-linear functional form $F\big(x(t-\tau)\big)$. The above conditions for local stability and non-oscillatory convergence can be applied to both these equations by using their respective functional forms. 
\subsection{Macky-Glass equation}
\label{sec:local_stability_MG}
The necessary and sufficient condition for local stability of the Mackey-Glass equation can be written as 
\begin{align}
 \frac{\gamma n \tau}{\beta}(\gamma-\beta)\sqrt{1 + 2\frac{\beta}{n(\gamma-\beta)}} < \cos^{-1}\bigg(\frac{\beta}{\beta + n(\gamma-\beta)}\bigg).\label{eq:necc-suff_MG}
\end{align}
If the above inequality is just violated, the system undergoes a Hopf bifurcation leading to the emergence of limit cycles whose period is given by
\begin{align}
 T =\, \frac{2\pi\beta}{\gamma n(\gamma-\beta)\sqrt{1+2\frac{\beta}{n(\gamma-\beta)}}}.\label{eq:Hopf_period_MG}
\end{align}
The sufficient condition, for local stability, is
\begin{align}
  \frac{\gamma\tau}{\beta} \Big(n(\beta-\gamma)-\beta\Big)< \frac{\pi}{2}.\label{eq:suff_MG}
 \end{align}
The underlying system equilibrates without any oscillations if and only if
\begin{align}
 \frac{\gamma\tau}{\beta}e^{\gamma\tau} \Big(n(\beta-\gamma)-\beta\Big)  < \frac{1}{e}\label{eq:non_oscill_MG}
\end{align}
is satisfied.
It can be observed that the above stability conditions depend on the model parameters as well as the time delay. In order to understand this interdependence better, we present some stability charts in Section~\ref{sec:stability_charts}.
 
\subsection{Lasota equation}
\label{sec:local_stability_LS}
 Similarly, for the Lasota equation, the necessary and sufficient condition for local stability is
 \begin{align}
  \gamma\tau\sqrt{(x^\ast-n)^2-1} < \cos^{-1}\bigg(\frac{1}{n-x^\ast}\bigg),\label{eq:necc-suff_LS}
 \end{align}
 where $x^\ast$ satisfies $\beta (x^\ast)^{(n-1)}e^{-x^\ast}=\gamma$. When the above condition is violated the system undergoes a Hopf bifurcation with period
 \begin{align}
  T = 2\pi\bigg/\bigg(\gamma\sqrt{(n-x^\ast)^2-1}\bigg).\label{eq:Hopf_period_LS}
 \end{align}
 The system can be stabilised if the following sufficient condition is satisfied
 \begin{align}
  \gamma\tau(x^\ast-n) < \frac{\pi}{2}.\label{eq:suff_LS}
 \end{align}
 The condition necessary and sufficient condition for non-oscillatory convergence is
 \begin{align}
  \gamma\tau(x^\ast-n) e^{\gamma\tau} < \frac{1}{e}.
 \end{align}
 
 \subsection{Stability charts}
 \label{sec:stability_charts}
 We now present some graphical representations of the stability conditions derived above. Such plots enable one to understand the trade-offs between various system parameters for maintaining system stability. 
 \begin{figure*}[t]
 \begin{center}
\captionsetup[subfigure]{labelformat=empty}
\psfrag{tcr}[l][c]{$\tau_{c}$}
  \psfrag{ts}[l][c]{$\tau_{suff}$}
  \psfrag{tn}[l][c]{$\tau_{noc}$}
   \psfrag{b}[c][b]{\hspace{35mm}Dependence parameter, $\beta$ (day$^{-1}$)}
  \psfrag{0}[b][b]{\scriptsize$0$}
   \psfrag{10}[b][b]{\scriptsize$10$}
     \psfrag{20}[b][b]{\scriptsize$20$}
   \psfrag{40}[b][b]{\scriptsize$40$}
     \psfrag{0.5}[b][b]{\scriptsize$0.5$}
   \psfrag{1}[b][b]{\scriptsize$1$}
     \psfrag{0.75}[b][b]{\scriptsize$0.75$}
       \psfrag{5}[b][b]{\scriptsize$5$}
     \psfrag{15}[b][b]{\scriptsize$15$}
   \psfrag{25}[b][b]{\scriptsize$25$}
     \psfrag{35}[b][b]{\scriptsize$35$}
  \psfrag{0}[b][b]{\scriptsize$0$}
   \psfrag{0.75}[b][b]{\scriptsize$0.75$}
     \psfrag{20}[b][b]{\scriptsize$20$}
   \psfrag{40}[b][b]{\scriptsize$40$}
     \psfrag{0.5}[b][b]{\scriptsize$0.5$}
   \psfrag{1}[b][b]{\scriptsize$1$}
     \psfrag{1.5}[b][b]{\scriptsize$1.5$}
       \psfrag{2}[b][b]{\scriptsize$2$}
     \psfrag{2.5}[b][b]{\scriptsize$2.5$}
   \psfrag{60}[b][b]{\scriptsize$60$}
     \psfrag{80}[b][b]{\scriptsize$80$}
      \psfrag{100}[b][b]{\scriptsize$100$}
           \psfrag{0.3}[b][b]{\scriptsize$0.3$}
   \psfrag{0.6}[b][b]{\scriptsize$0.6$}
     \psfrag{0.9}[b][b]{\scriptsize$0.9$}
      \psfrag{1.2}[b][b]{\scriptsize$1.2$}
      \begin{tabular}{c c c}
     \toprule
       Mackey-Glass equation & & Lasota equation\\    
        \cmidrule(r){1-1}\cmidrule(lr){3-3}
       \subfloat[]{
       \psfrag{t}[b][b]{\hspace{1mm}Feedback delay, $\tau$ (days)}
\includegraphics[width=1.6in,height=2.4in,angle=270]{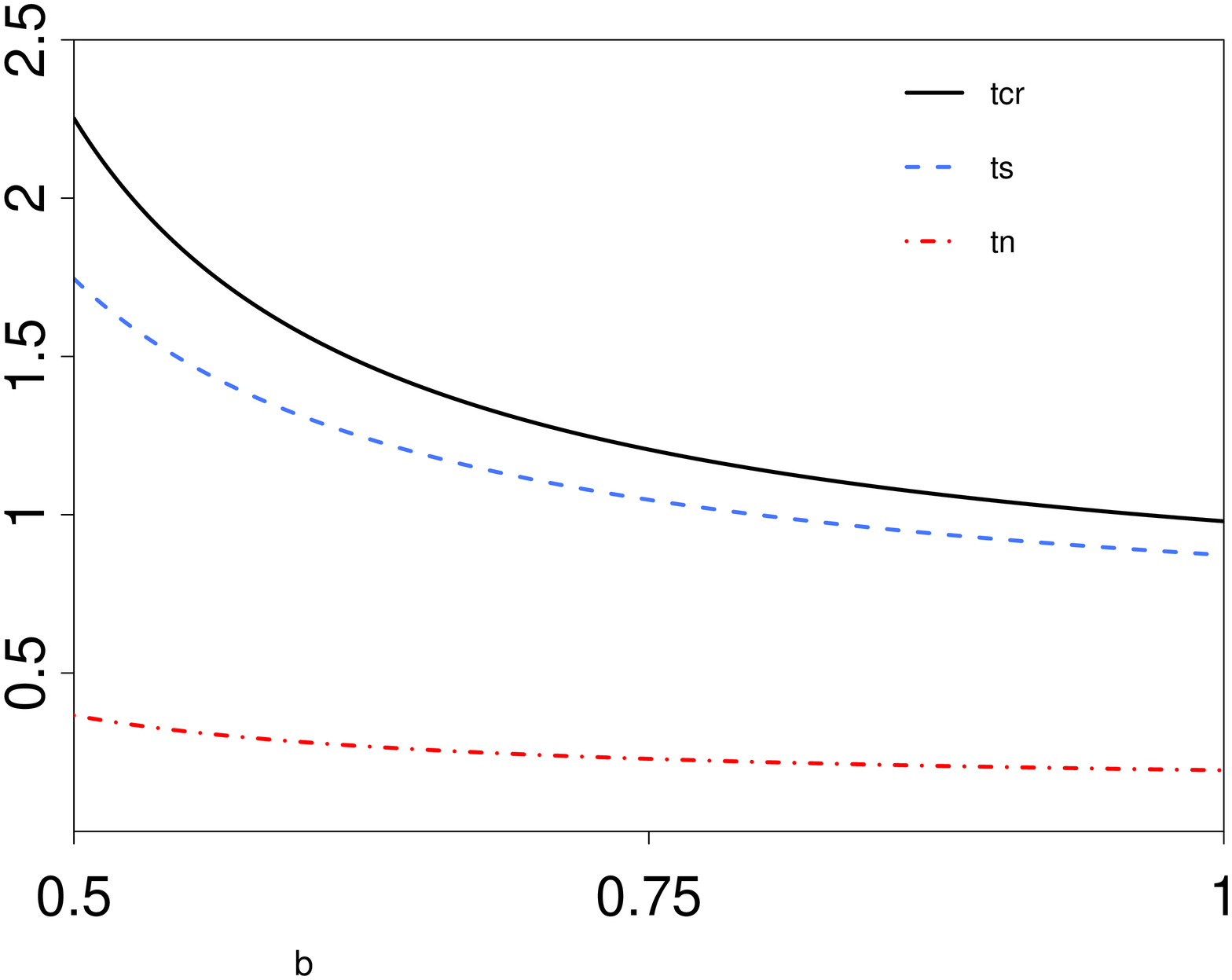}
}
& &
     \subfloat[]{
     \psfrag{t}[b][b]{\hspace{2.5mm}Feedback delay, $\tau$ (days)}
\includegraphics[width=1.6in,height=2.4in,angle=270]{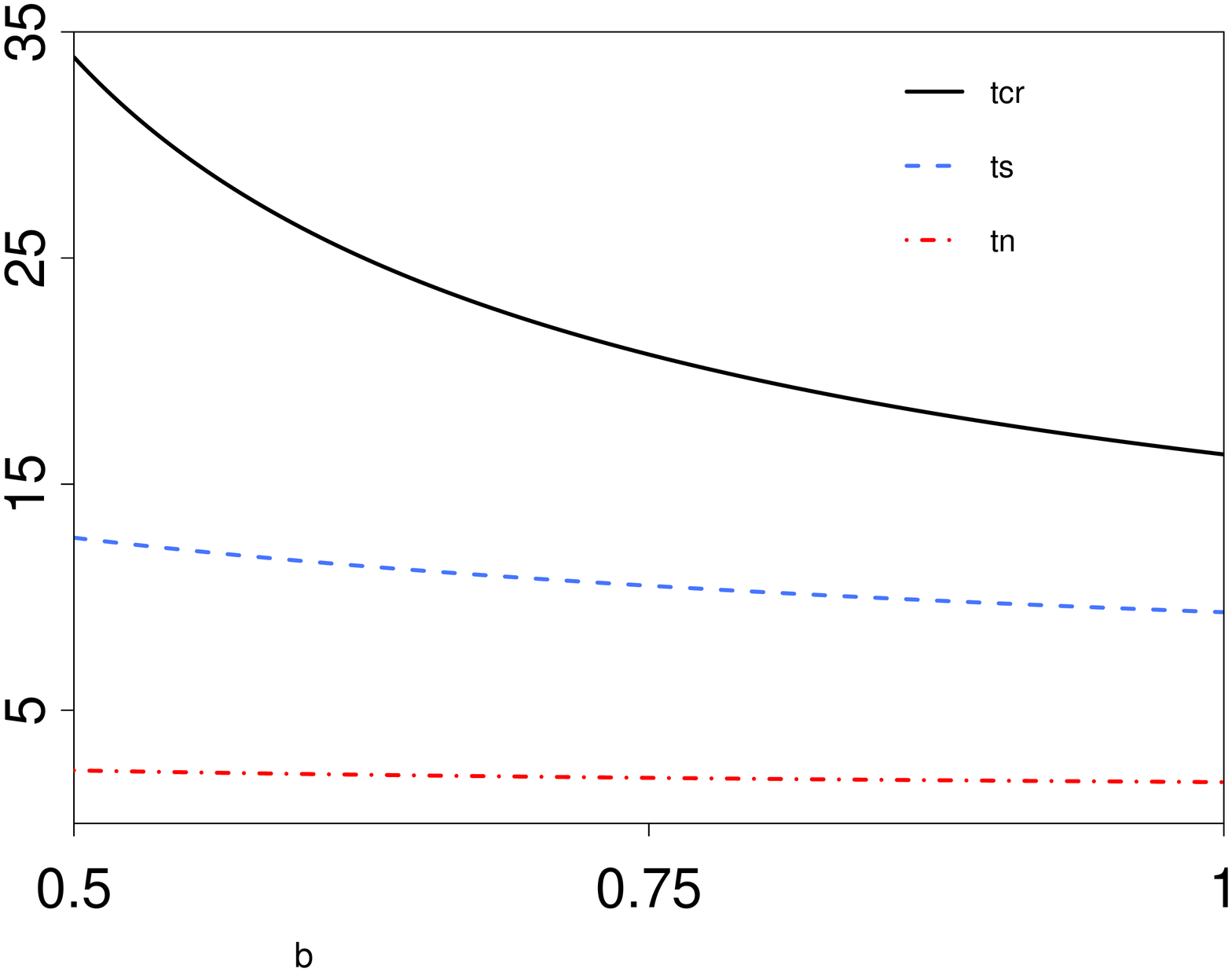}
}
\\
\subfloat[]{
\psfrag{t}[b][b]{\hspace{1mm}Feedback delay, $\tau$ (days)}
\psfrag{n}[c][b]{\hspace{5mm}Non-linearity parameter, $n$}
\includegraphics[width=1.6in,height=2.4in,angle=270]{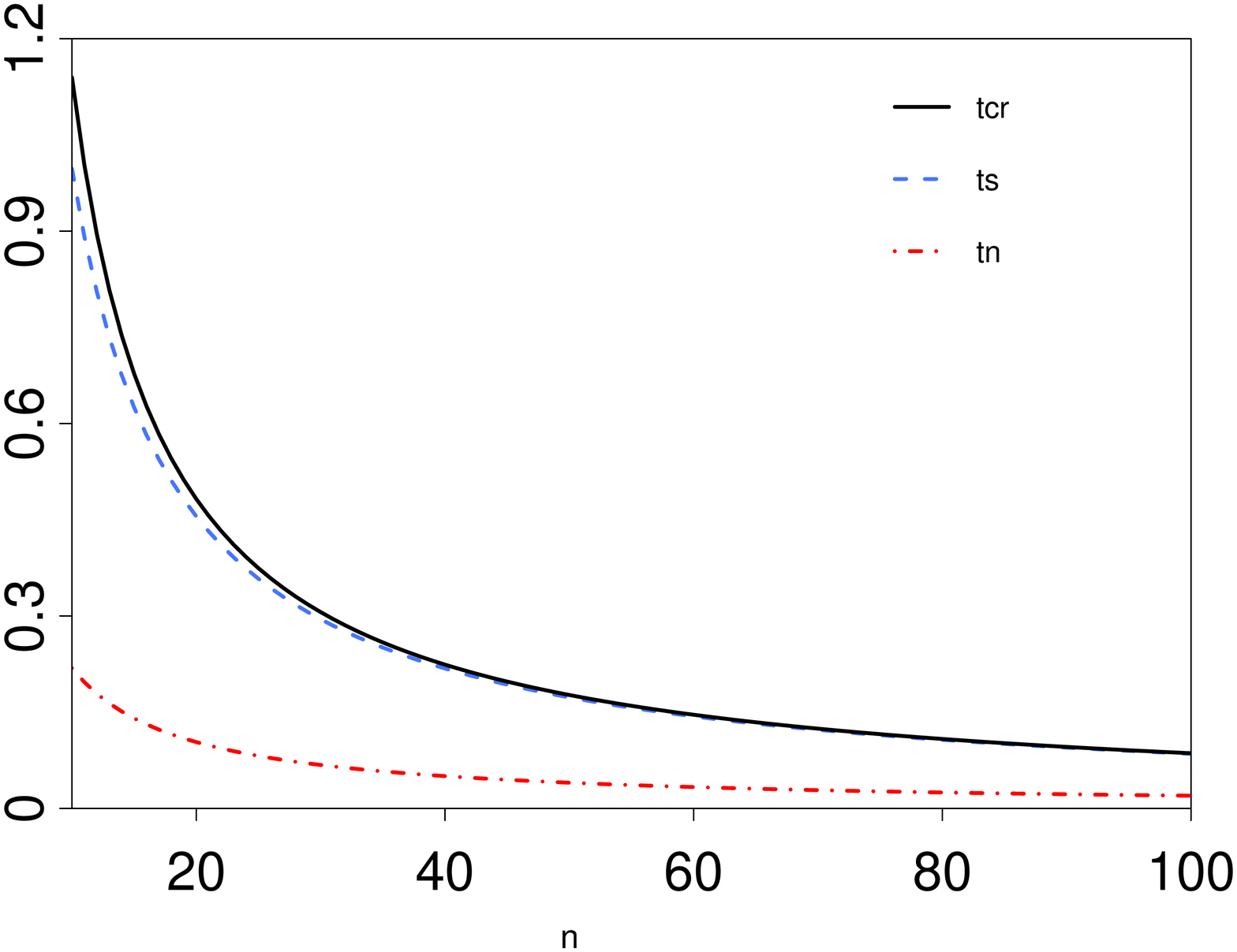}
}
& &
\subfloat[]{
\psfrag{t}[b][b]{\hspace{1mm}Feedback delay, $\tau$ (days)}
\psfrag{n}[c][b]{\hspace{10mm}Non-linearity parameter, $n$}
\includegraphics[width=1.6in,height=2.4in,angle=270]{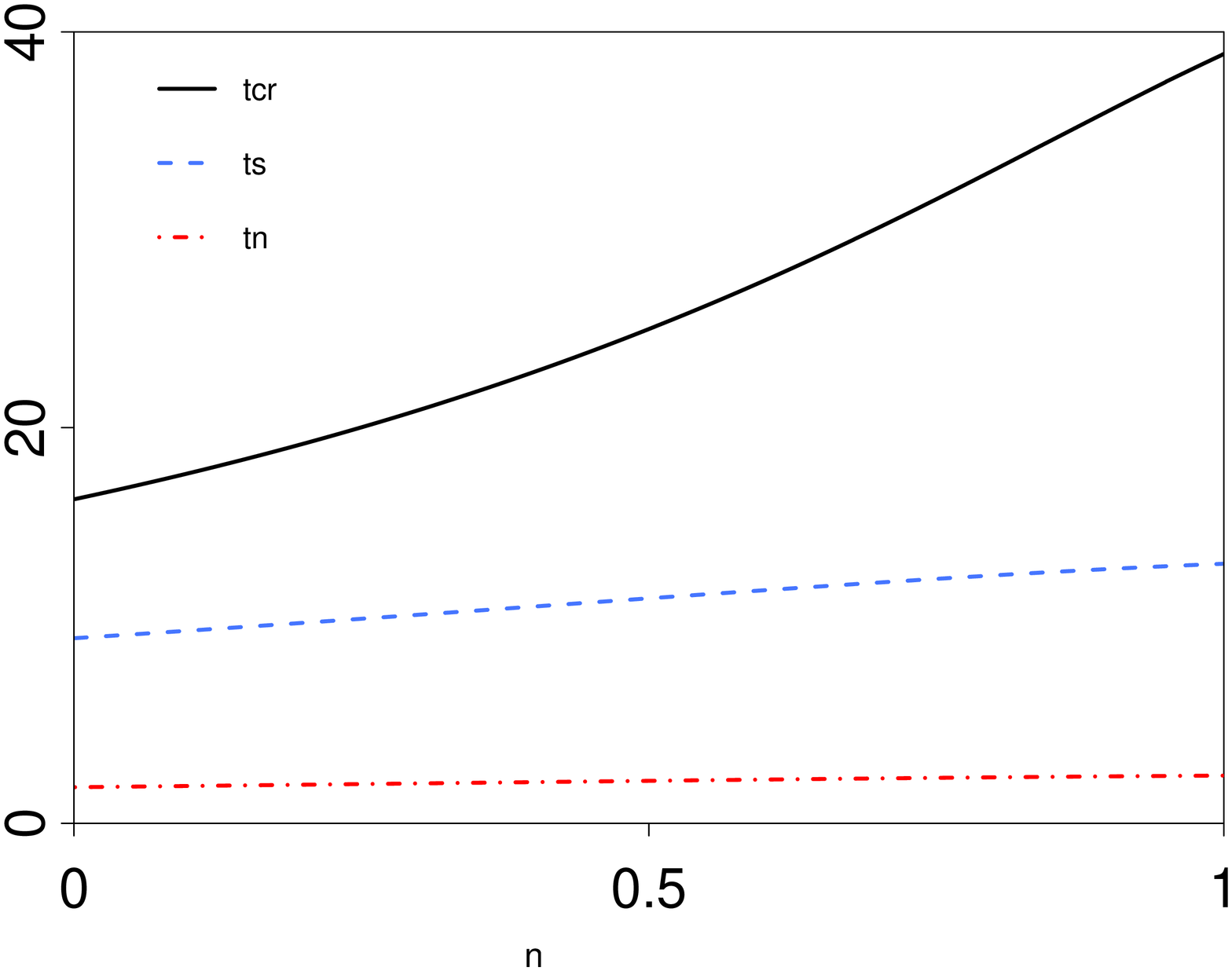}
}\\
\bottomrule
      \end{tabular}
      \end{center}
      \caption{Local stability charts for the Mackey-Glass and Lasota equations. There exist trade-offs between the choice of parameters $\beta$ and $n$ and the feedback delay, for both models.}\label{fig:stability_plots}
      \end{figure*}

 \subsubsection{Mackey-Glass equation}
 We begin by trying to understand the trade-off between the parameter $\beta$\textemdash which captures the dependence of the rate of change of concentration of mature blood cells $\dot{x}(t)$ on the number of mature cells in the blood stream\textemdash and the feedback delay $\tau$. For this, we fix the other parameters as follows: $\gamma = 0.3, n = 10$, and vary $\beta$ in the range $[0.5,1]$. For each value of $\beta$ in this range, we compute the values of $\tau_c$, $\tau_{suff}$ and $\tau_{noc}$, which define the boundaries of the necessary and sufficient condition~\eqref{eq:necc-suff_MG}, sufficient condition~\eqref{eq:suff_MG} and the condition~\eqref{eq:non_oscill_MG} for non-oscillatory convergence. This plot is presented in Fig.~\ref{fig:stability_plots}. Recall that, the system undergoes a Hopf bifurcation when $\tau = \tau_c$. This implies that when $\tau=\tau_c$, the system would lose stability leading to the emergence of limit cycles in the concentration of blood cells. Observe that, when the feedback delay $\tau$ increases, the parameter $\beta$ would have to necessarily reduce in order to maintain system stability. Also observe that the sufficient condition $\tau < \tau_{suff}$ is not very conservative, and may hence yield design guidelines. The parameter $\beta$ may be varied as per the sufficient condition to ensure system stability. 
 
 \begin{figure*}[t]
\begin{center}
 \psfrag{0}[b][b]{\scriptsize$0$}
 \psfrag{1.0}[b][b]{\scriptsize$1.0$}
   \psfrag{2.5}[b][b]{\scriptsize$2.5$}
     \psfrag{5}[b][b]{\scriptsize$5$}
   \psfrag{7.5}[b][b]{\scriptsize$7.5$}
     \psfrag{10}[b][b]{\scriptsize$10$}
   \psfrag{1}[b][b]{\scriptsize$1$}
     \psfrag{0.8}[b][b]{\scriptsize$0.8$}
       \psfrag{0.6}[b][b]{\scriptsize$0.6$}
     \psfrag{1.2}[b][b]{\scriptsize$1.2$}
   \psfrag{1.4}[b][b]{\scriptsize$1.4$}
        \psfrag{1.6}[b][b]{\scriptsize$1.6$}
       \psfrag{2}[b][b]{\scriptsize$2$}
     \psfrag{2.4}[b][b]{\scriptsize$2.4$}
   \psfrag{2.8}[b][b]{\scriptsize$2.8$}
     \psfrag{3.2}[b][b]{\scriptsize$3.2$}
   \psfrag{tcr}[c][b]{\hspace{7mm}\scriptsize{$\tau > \tau_c$}}
   \psfrag{ts}[c][b]{\hspace{21mm}\scriptsize{$\tau_{noc} < \tau < \tau_{suff}$}}
   \psfrag{tn}[c][b]{\hspace{10mm}\scriptsize{$\tau < \tau_{noc}$}}
   \psfrag{x}[b][b]{\hspace{10mm}$x(t)$ (cells/kg)}
   \subfloat[Mackey-Glass equation]{
   \psfrag{t}[b][b]{\hspace{15mm}Time (days)}
\includegraphics[width=1.6in,height=2.4in,angle=270]{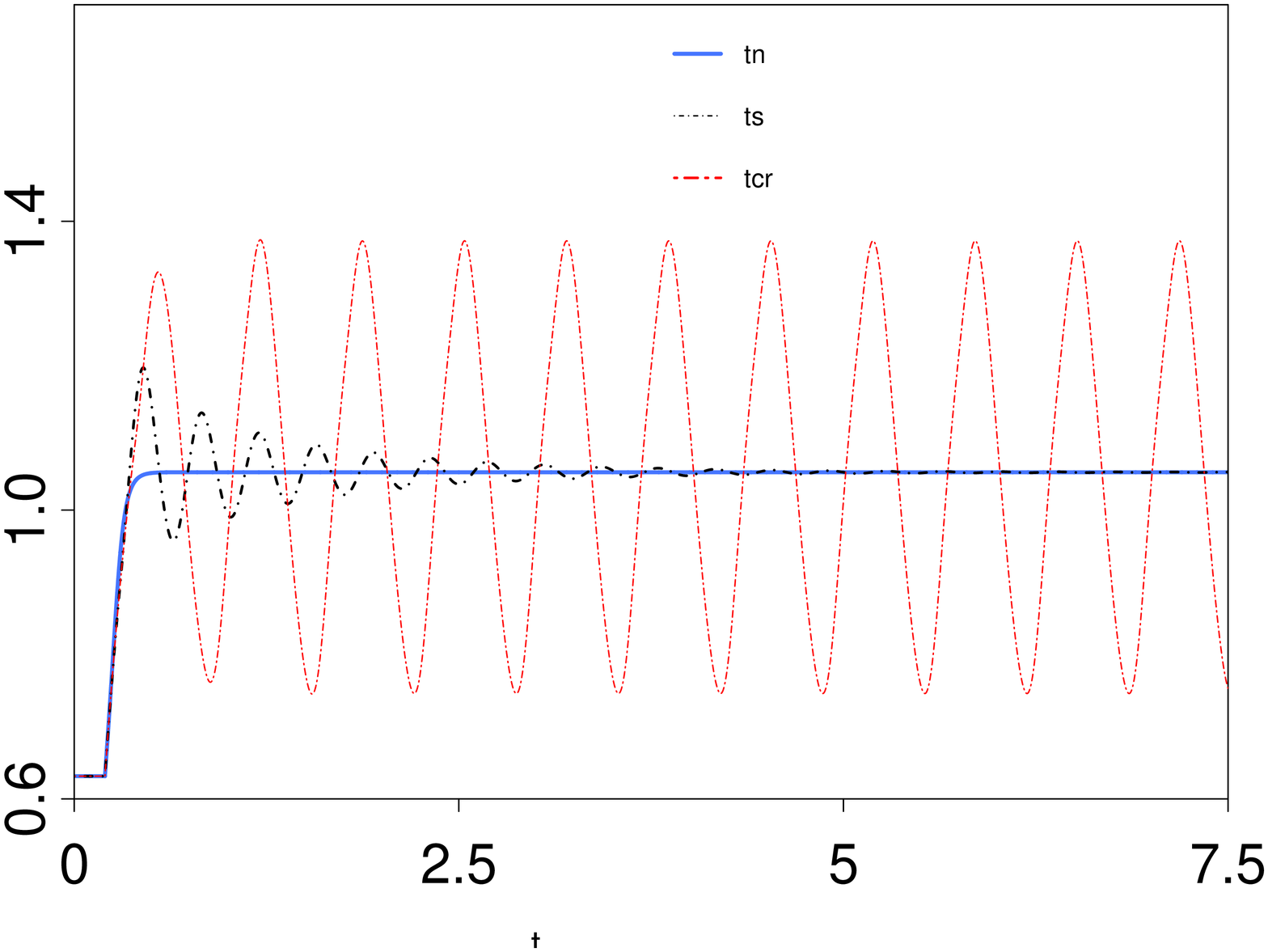}
\label{fig:time_response_MG}
}
\qquad
   \subfloat[Lasota equation]{
   \psfrag{t}[b][b]{\hspace{8mm}Time (days)}
\includegraphics[width=1.6in,height=2.4in,angle=270]{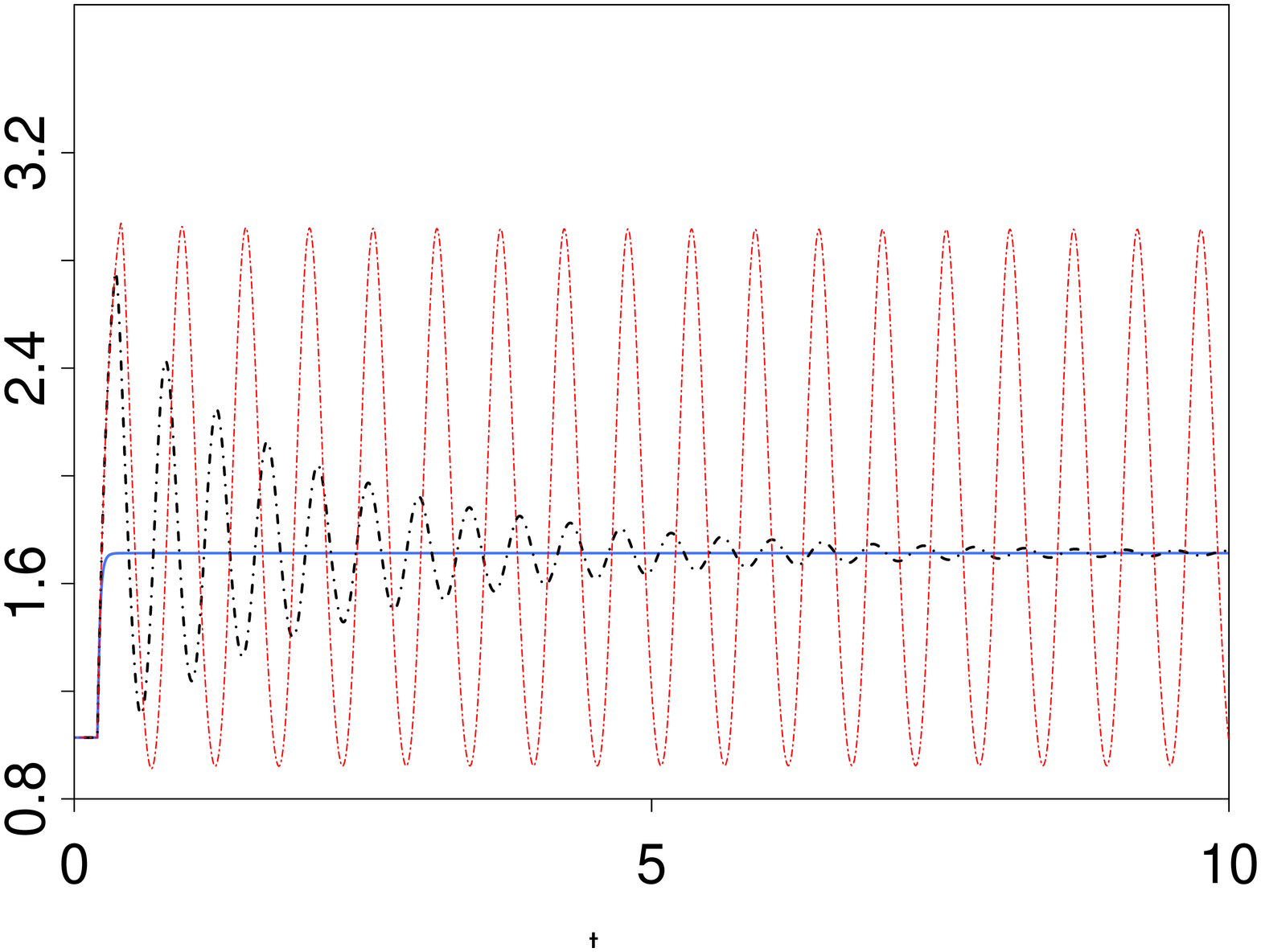}\label{fig:time_response_LS}
}
\caption{Time response of (a) Mackey-Glass equation, (b) Lasota equation. For $\tau < \tau_{noc}$, the system reaches the equilibrium without any oscillations, for $ \tau_{noc}<\tau < \tau_{suff}$, the system equilibrates via non-linear oscillations. When $\tau > \tau_{cr}$, the system is unstable, and we observe the emergence of a limit cycle.}
\end{center}
\end{figure*}
 
 We then constructed a similar plot to understand the relationship between the feedback delay $\tau$ and the non-linearity parameter $n$ (shown in Fig.~\ref{fig:stability_plots}). For this plot, the parameter $\beta = 0.8$ and the parameter $n$ is varied in the range $[10,100]$.
 A similar trend is observed in this plot as well: the non-linearity parameter $n$ would have to reduce as feedback delay $\tau$ increases, in order to ensure system stability.
 
Time response curves for the Mackey-Glass equation plotted in Fig.~\ref{fig:time_response_MG} depicts the variation in the concentration of blood cells with respect to time. Observe that when the feedback delay is less than the value $\tau_{noc}$, the system is in the region of non-oscillatory convergence. The solution converges to the equilibrium without any oscillations. When the feedback delay is such that the sufficient condition~\eqref{eq:suff_MG} is satisfied, \emph{i.e.,} $\tau < \tau_{suff}$, the system is stable, and the solution exhibits damped oscillations followed by convergence to equilibrium. The feedback delay is more than the critical time delay $\tau > \tau_c$, the necessary and sufficient condition is violated and the system transits into instability via a Hopf bifurcation. This is marked by the emergence of un-damped oscillations in the system dynamics, which manifest as oscillations in the concentration of the blood cells. 

\begin{figure*}[t]
 \begin{center}
\captionsetup[subfigure]{labelformat=empty}
\psfrag{tcr}[l][c]{$\tau_{c}$}
  \psfrag{ts}[l][c]{$\tau_{suff}$}
  \psfrag{tn}[l][c]{$\tau_{noc}$}
   \psfrag{b}[c][b]{\hspace{35mm}Dependence parameter, $\beta$ (day$^{-1}$)}
   \psfrag{g}[c][b]{\hspace{31mm}Destruction rate, $\gamma$ (day$^{-1}$)}
   \psfrag{3.5}[b][b]{\hspace{2mm}$3.5$}
   \psfrag{4.5}[b][b]{\scriptsize$4.5$}
    \psfrag{5.5}[b][b]{\scriptsize$5.5$}
     \psfrag{6.5}[b][b]{\scriptsize$6.5$}
     \psfrag{7.5}[b][b]{\scriptsize$7.5$}
   \psfrag{8.5}[b][b]{\scriptsize$8.5$}
  \psfrag{0}[b][b]{\scriptsize$0$}
   \psfrag{0.1}[b][b]{\scriptsize$0.1$}
   \psfrag{10}[b][b]{\scriptsize$10$}
     \psfrag{20}[b][b]{\scriptsize$20$}
   \psfrag{40}[b][b]{\scriptsize$40$}
     \psfrag{0.5}[b][b]{\scriptsize$0.5$}
   \psfrag{1}[b][b]{\scriptsize$1$}
     \psfrag{0.75}[b][b]{\scriptsize$0.75$}
       \psfrag{5}[b][b]{\scriptsize$5$}
       \psfrag{85}[b][b]{\scriptsize$85$}
     \psfrag{65}[b][b]{\scriptsize$65$}
       \psfrag{45}[b][b]{\scriptsize$45$}
     \psfrag{15}[b][b]{\scriptsize$15$}
   \psfrag{25}[b][b]{\scriptsize$25$}
     \psfrag{35}[b][b]{\scriptsize$35$}
  \psfrag{0}[b][b]{\scriptsize$0$}
   \psfrag{0.75}[b][b]{\scriptsize$0.75$}
     \psfrag{20}[b][b]{\scriptsize$20$}
   \psfrag{40}[b][b]{\scriptsize$40$}
     \psfrag{0.5}[b][b]{\scriptsize$0.5$}
   \psfrag{1}[b][b]{\scriptsize$1$}
     \psfrag{1.5}[b][b]{\scriptsize$1.5$}
     \psfrag{120}[b][b]{\scriptsize$120$}
     \psfrag{180}[b][b]{\scriptsize$180$}
       \psfrag{2}[b][b]{\scriptsize$2$}
     \psfrag{2.5}[b][b]{\scriptsize$2.5$}
   \psfrag{60}[b][b]{\scriptsize$60$}
     \psfrag{80}[b][b]{\scriptsize$80$}
      \psfrag{100}[b][b]{\scriptsize$100$}
           \psfrag{0.3}[b][b]{\scriptsize$0.3$}
   \psfrag{0.6}[b][b]{\scriptsize$0.6$}
     \psfrag{0.9}[b][b]{\scriptsize$0.9$}
      \psfrag{1.2}[b][b]{\scriptsize$1.2$}
      \begin{tabular}{c c c}
     \toprule
       Mackey-Glass equation & & Lasota equation\\    
        \cmidrule(r){1-1}\cmidrule(lr){3-3}
       \subfloat[]{
       \psfrag{t}[b][b]{\hspace{25mm}Time period, $T$ (days)}
\includegraphics[width=1.6in,height=2.4in,angle=270]{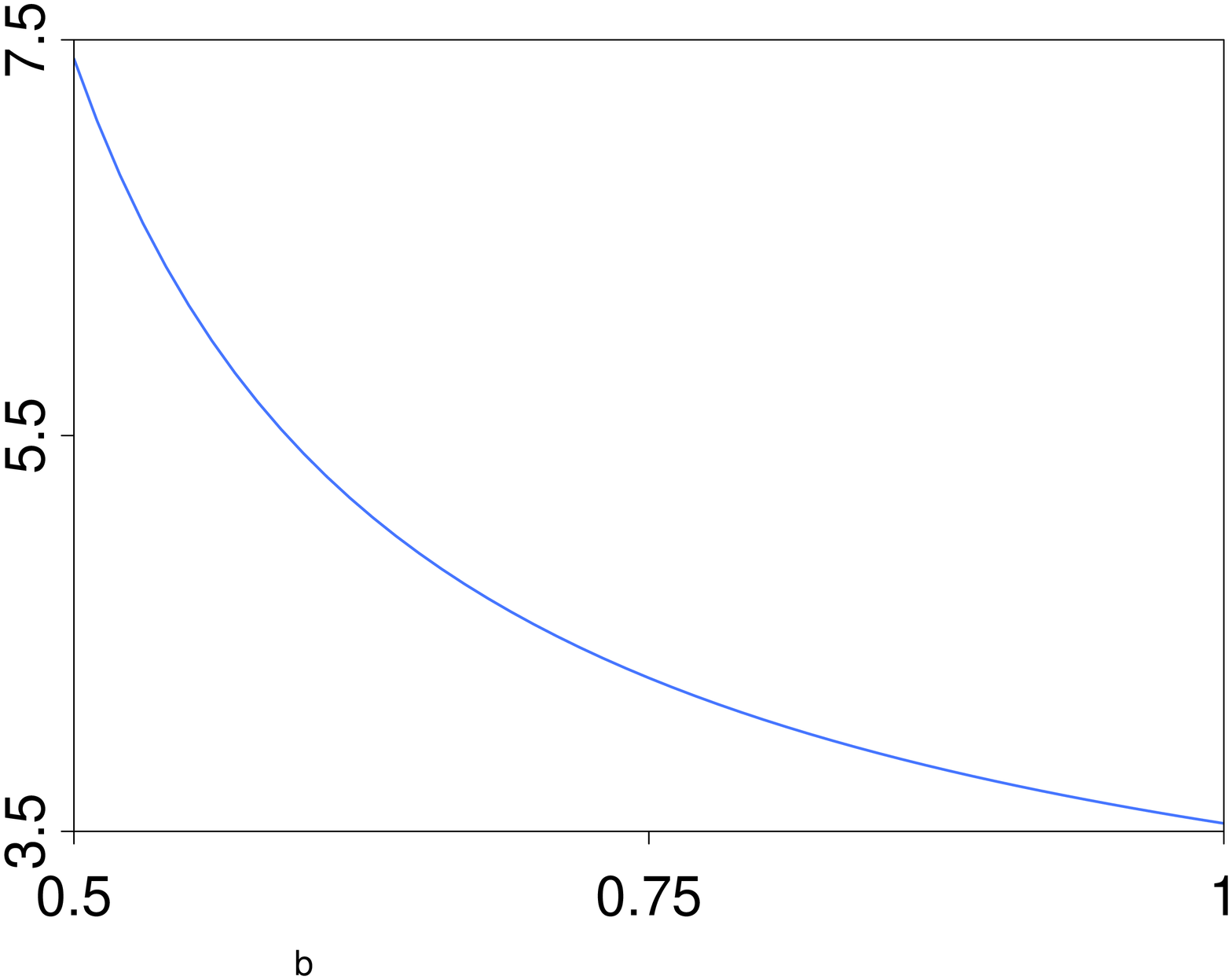}
}
& &
     \subfloat[]{
     \psfrag{t}[b][b]{\hspace{25mm}Time period, $T$ (days)}
\includegraphics[width=1.6in,height=2.4in,angle=270]{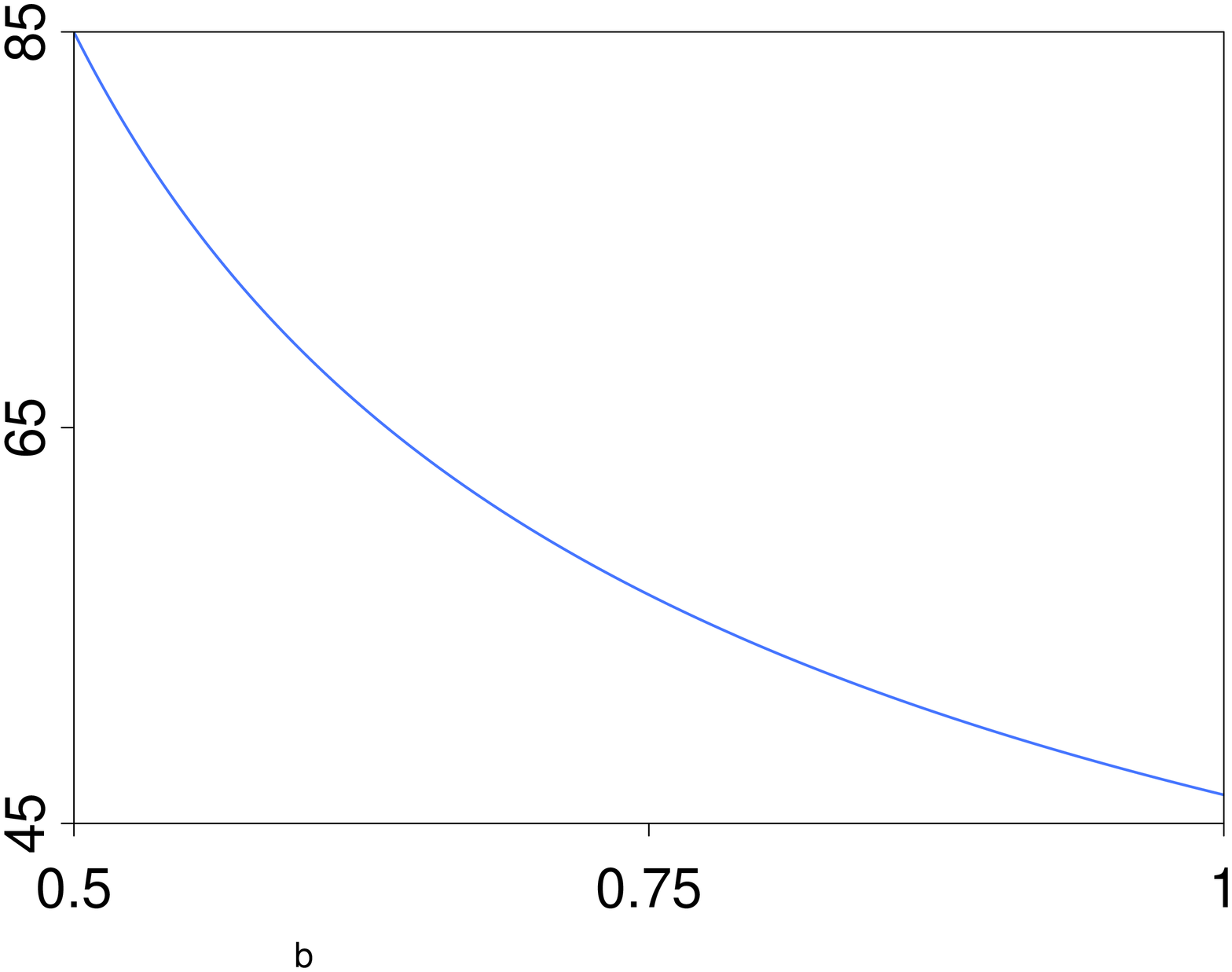}
}
\\
\subfloat[]{
\psfrag{t}[b][b]{\hspace{25mm}Time period, $T$ (days)}
\includegraphics[width=1.6in,height=2.4in,angle=270]{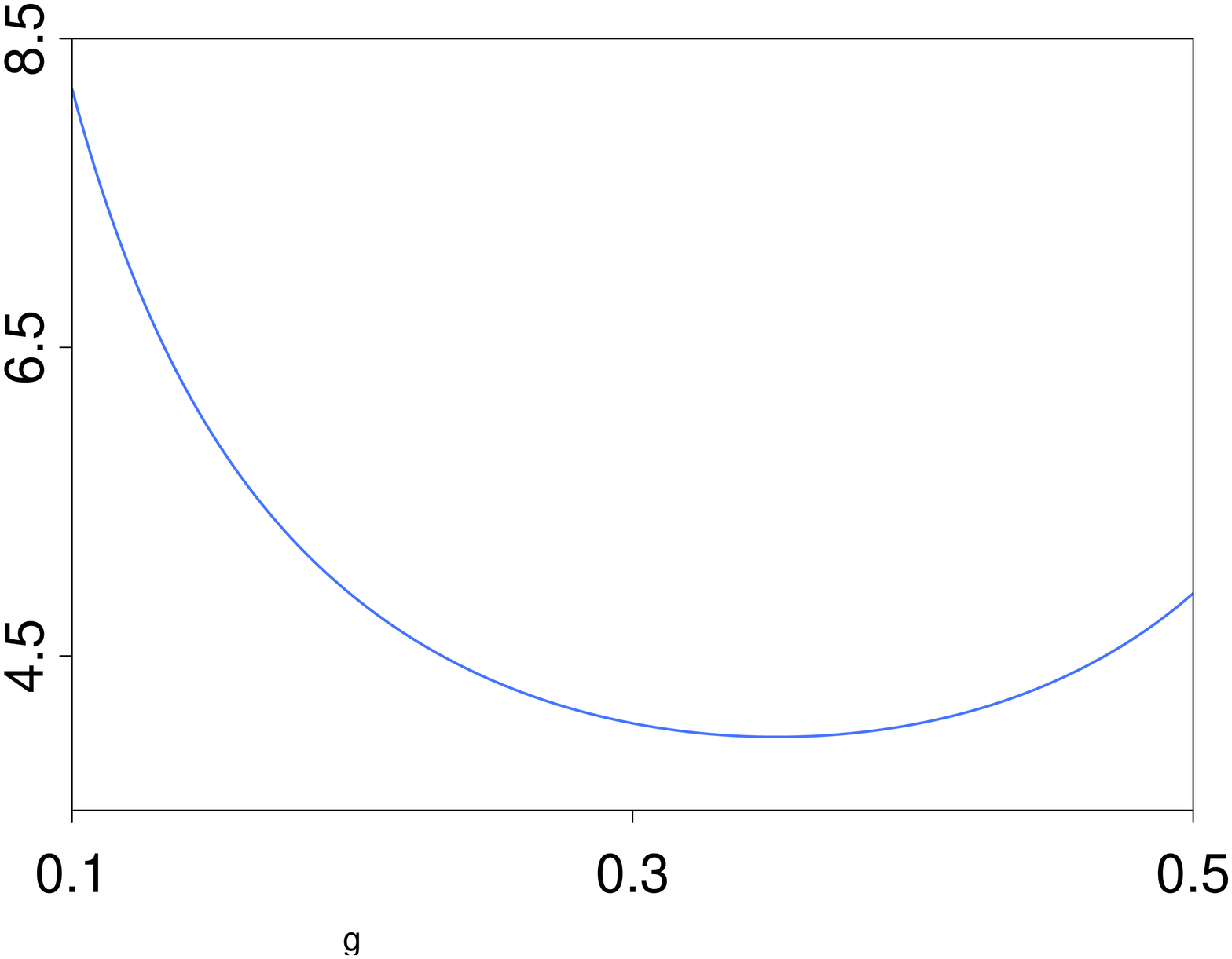}
}
& &
\subfloat[]{
\psfrag{t}[b][b]{\hspace{25mm}Time period, $T$ (days)}
\includegraphics[width=1.6in,height=2.4in,angle=270]{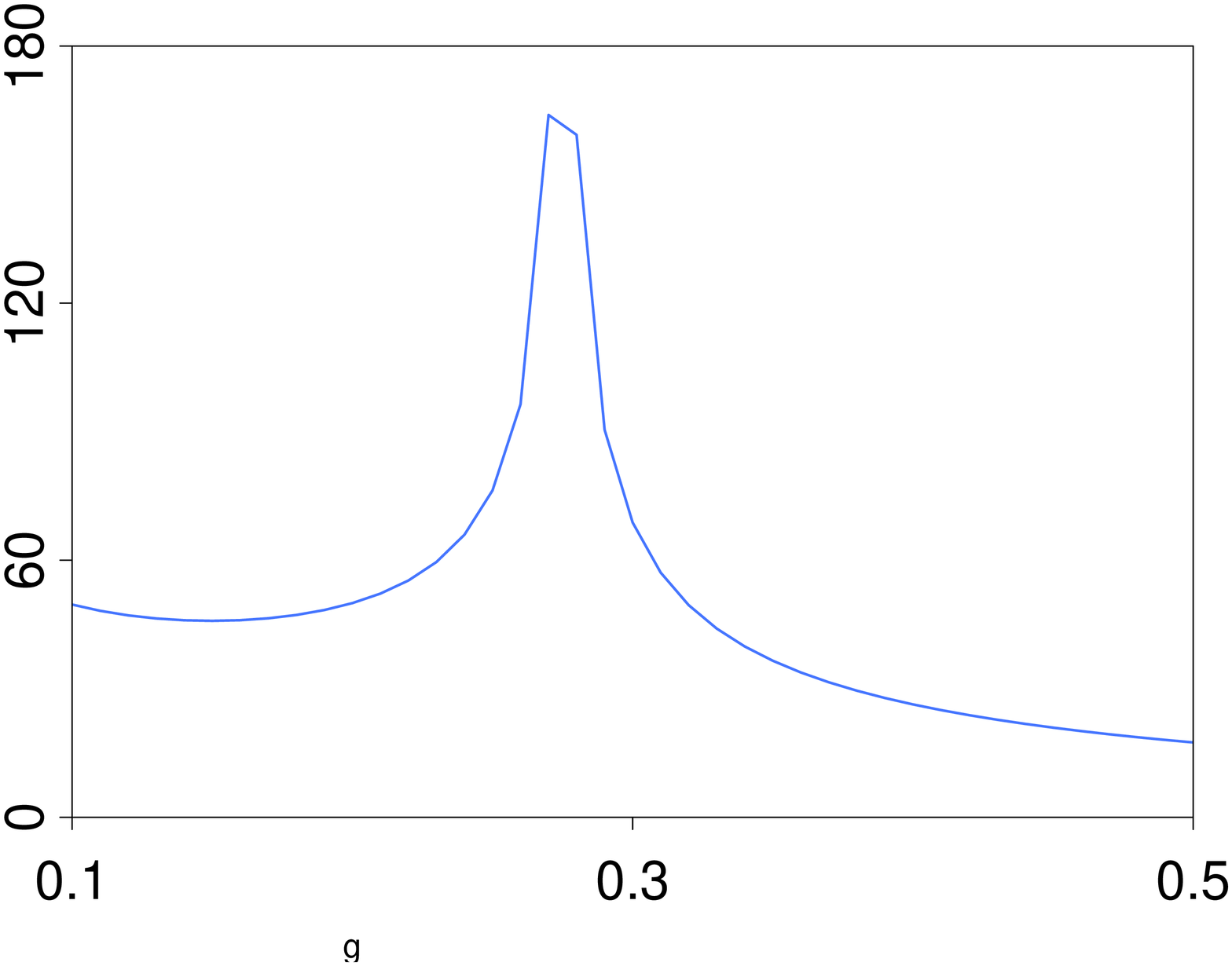}
}\\
\bottomrule
      \end{tabular}
      \end{center}
      \caption{Dependence of time period of the limit cycles on various model parameters in Mackey-Glass and Lasota equations. Observe that the time period reduces as the parameter $\beta$ increases. However, the dependence on the destruction rate $\gamma$ is qualitatively different for the two models.}\label{fig:time_period_dependence}
      \end{figure*}

Having observed that loss of local stability leads to limit cycles in the concentration of the blood cells, it may be desirable to examine the time period of these oscillations, and understand the dependence of this time period on various system parameters. Therefore, we plotted the time period, given by equation~\eqref{eq:Hopf_period_MG} as a function of the parameter $\beta$ and the cell destruction rate $\gamma$. These plots are presented in Fig.~\ref{fig:time_period_dependence}. It can be observed that the time period reduces with an increase in the parameter $\beta$. This signifies that, as the dependence parameter $\beta$ increases, the blood cell concentration oscillates with higher frequency, which could indicate increasing severity of a dynamical disease. However, as the destruction rate $\gamma$ increases, the time period reduces until it attains a minimum value, after which it increases. 

 \subsubsection{Lasota equation}
 We also construct stability charts for the Lasota equation. For these plots, we fix the parameters as $\gamma = 0.1, n = 0.1$, and then vary the parameter $\beta$ in the range $[0.5,1]$. We obtain the trade-off between the feedback delay and the parameter $\beta$ as shown in Fig.~\ref{fig:stability_plots}. The qualitative observations are the same as in the case of the Mackey-Glass equation, \emph{i.e.}, in order to maintain local stability, the parameter $\beta$ would have to reduce as the feedback delay $\tau$ increases. 
 We then fix $\beta = 0.9$ and vary the non-linearity parameter $n$ in the range $[0.001,1]$. This plot shows that as the feedback delay increases, the parameter $n$ would also have to increase in order to maintain stability. This is qualitatively different from the trade-off observed in case of the Mackey-Glass equation. The time response plots for the Lasota equation are presented in Fig.~\ref{fig:time_response_LS}. The three qualitative behaviours of the system are shown for various values of the feedback delay. The dependence of the time period of limit cycles on the parameter $\beta$ and the destruction rate $\gamma$ is 
shown in Fig.~\ref{fig:time_period_dependence}.
 
 \subsection{Discussion}
  So far, we have characterised the system stability by finding the bounds on the system parameters and the feedback delay for stability. This enables us to design and configure system parameters such that the concentration of the blood cells converges to a desired level. However, in order to avoid dynamical diseases, it is also important to ensure that this convergence is achieved within a specified time frame. To that end, it would be desirable to study the convergence characteristics of these systems. In the next section, we define closed-form expressions for the rate of convergence of the solutions of both models. 
 
 \section{Rate of convergence}
\label{sec:roc}
In this section, we study the convergence characteristics of the two models that we consider. We first present the analysis for the general first-order, linear, time-delayed system~\eqref{eq:general-lin}. Then, we interpret these results for the Mackey-Glass and the Lasota equation. 

The convergence characteristics of the solutions, about the stable equilibrium, can be studied with respect to the feedback delay $\tau$. 
To do so, it is sufficient to solve the characteristic equation of \eqref{eq:general-char}, whose roots determine the convergence characteristics of solutions of \eqref{eq:general-lin}.
We closely follow the style of analysis outlined in~\cite{Brauer_79}. To analyse the dependence on the feedback delay $\tau$, we require that $b\neq0$. 
Using the transformation $\lambda\tau=z$, $-a\tau=p$ and $-b\tau=q$ in the characteristic equation~\eqref{eq:general-char}, we get
\begin{align}
 (p-z)e^{z}+q=0.\label{eq:mce}
\end{align}
If $-\alpha<0$ is the real part of a root of \eqref{eq:mce}, then \eqref{eq:general-lin} has a solution of the form $e^{-(\alpha/\tau)t}$, which is a decaying function of $t$. If every root of the characteristic equation \eqref{eq:mce} lies in the open left half of the Argand plane, then the system is asymptotically stable. Here, $\alpha/\tau$ is the convergence rate, the rate at which the stable system equilibrates. The inverse of the rate of convergence, $\tau/\alpha$ is the characteristic return time, where $-\alpha$ is the largest of the real parts of all the roots of \eqref{eq:mce}. A necessary and sufficient condition for all roots of \eqref{eq:mce} to be in the left half of the Argand plane is~\cite{Hayes_50}
\begin{align}
 p<1, \,\,\,\,\,  p<- q< \frac{u_1}{\sin (u_1)},\label{eq:sc}
\end{align}
where $u_1$ is the solution of the equation 
\begin{align}
u = p\tan (u), \label{eq:sc1}
\end{align}
in $u\in(0,\pi)$, with $u_1=\pi/2$ if $p=0$.
Using the change of variable $z=\chi-\sigma\tau$, where $\sigma = \alpha/\tau$, characteristic equation~\eqref{eq:mce} can be transformed to
\begin{align}
 \big((p+\sigma\tau)-\chi\big)e^{\chi}+qe^{\sigma\tau}=0.\label{eq:cez}
\end{align}
Let $\sigma$ be the supremum of the solution of \eqref{eq:cez} over $(0,\infty)$ which guarantees that all roots of the transformed characteristic equation \eqref{eq:cez} lie in the left half of the Argand plane. Then, $\sigma$ is the rate of convergence to the equilibrium of \eqref{eq:general-lin}.
For the transformed characteristic equation \eqref{eq:cez}, the set of inequalities in~\eqref{eq:sc} can be restated as
\begin{align}
 p+\sigma\tau = (- a + \sigma)\tau &< 1,\label{eq:msc1}\\
 p+\sigma\tau = (- a + \sigma)\tau &< - q e^{\sigma\tau} = b\tau e^{\sigma\tau},\label{eq:msc2}\\
 - q e^{\sigma\tau} = b\tau e^{\sigma\tau} &< \frac{u_2}{\sin (u_2)},\label{eq:msc3}
\end{align}
where $u_2$ is the solution of the equation 
\begin{align}
u =(- a + \sigma)\tau \tan (u),\label{eq:msc4} 
\end{align}
in  the range $(0, \pi).$ If $ - a + \sigma = 0 ,$ then we have $u_2=\pi/2.$ The supremum of the solution of \eqref{eq:cez} over $(0,\infty)$, $\sigma$ satisfies the inequalities~\eqref{eq:msc1}-\eqref{eq:msc3}.
Equation \eqref{eq:msc2} can be re-written as
\begin{align}
 (- a + \sigma)\tau e^{(a - \sigma)\tau} < b\tau e^ {a\tau}.\label{eq:msc21}
\end{align}
Consider the function,
\begin{align*}
 g(u) = \frac{u}{\sin (u)} e^{- u/ \tan (u)}.
\end{align*}
For $u \in (0,\pi)$, $u/\sin (u)$ is an increasing function of $u$, $u/\tan (u)$ is a decreasing function of $u$. As a result, $g(u)$ is an increasing function of $u$. One may observe that $g(0)=1/e$, $g(\pi/2) = \pi/2$, $\lim_{u\to\pi} g(u) = \infty$.
Using \eqref{eq:msc4}, the inequality \eqref{eq:msc3} can be written as
\begin{align}
 b\tau e^ {a\tau} < g(u_2).\label{eq:msc31}
\end{align}
It can be observed from \eqref{eq:msc4} that $u_2$ is a decreasing function of $\sigma$. To obtain the maximum $\sigma$ satisfying the inequality \eqref{eq:msc31}, we need to solve its corresponding equality.
Similarly, $(- a + \sigma)\tau$ from \eqref{eq:msc1} and $(- a + \sigma)\tau e^{(a - \sigma)\tau}$ from \eqref{eq:msc21} are increasing functions of $\sigma$, and to obtain the maximum $\sigma$ satisfying the inequalities \eqref{eq:msc1}, \eqref{eq:msc21}, we need to solve the corresponding equalities. If there is no solution $\sigma$ satisfying the equalities corresponding to the inequalities \eqref{eq:msc1}, \eqref{eq:msc21} and \eqref{eq:msc31}, there is  no restriction on $\sigma$. These results can be summarised as follows.

Let $\sigma_1,~\sigma_2,~\sigma_3$ be the solutions of
\begin{align}
(- a + \sigma)\tau &= 1,\label{eq:msc5}\\
(- a + \sigma)\tau e^{(a - \sigma)\tau} & = b\tau e^ {a\tau},\label{eq:msc6}\\
g(u_2) &= b\tau e^ {a\tau},\notag\\
u_2 &= (- a + \sigma)\tau \tan (u_2),\label{eq:msc7}
\end{align}
respectively. We get $\sigma_i = \infty$, for $i={1,~2,~3}$ if the corresponding equality has no solution. Then the rate of convergence of the solutions of~\eqref{eq:general-lin} is given by
\begin{align*}
 \sigma = \min [\sigma_1, \sigma_2, \sigma_3].
\end{align*}

\begin{figure*}
 \begin{center}
 \psfrag{t}[b][b]{Feedback delay, $\tau$ (days)}
 \psfrag{0.0000}[b][b]{\scriptsize$0$}
 \psfrag{1.4273}[b][b]{\scriptsize$\tau^{\ast}$}
 \psfrag{3.0000}[b][b]{\scriptsize$3$}
 \psfrag{4.0000}[b][b]{\scriptsize$4$}
 \psfrag{5.0000}[b][b]{\scriptsize$5$}
 \psfrag{0.45}[b][b]{\scriptsize$0.45$}
 \psfrag{0.60}[b][b]{\scriptsize$0.6$}
 \psfrag{0.80}[b][b]{\scriptsize$0.8$}
 \psfrag{1.00}[b][b]{\scriptsize$1$}
 \psfrag{0.2189}[b][b]{\scriptsize$\tau^{\ast}$}
 \psfrag{0.5000}[b][b]{\scriptsize$0.5$}
 \psfrag{1.8}[b][b]{\scriptsize$1.8$}
 \psfrag{3.0}[b][b]{\scriptsize$3$}
 \psfrag{4.0}[b][b]{\scriptsize$4$}
 \psfrag{5.0}[b][b]{\scriptsize$5$}
  \subfloat[Mackey-Glass equation]{
  \psfrag{a}[b][b]{\hspace{-2mm}Convergence rate ($\text{days}^{-1}$)}
  \includegraphics[width=1.6in,height=2.2in,angle=270]{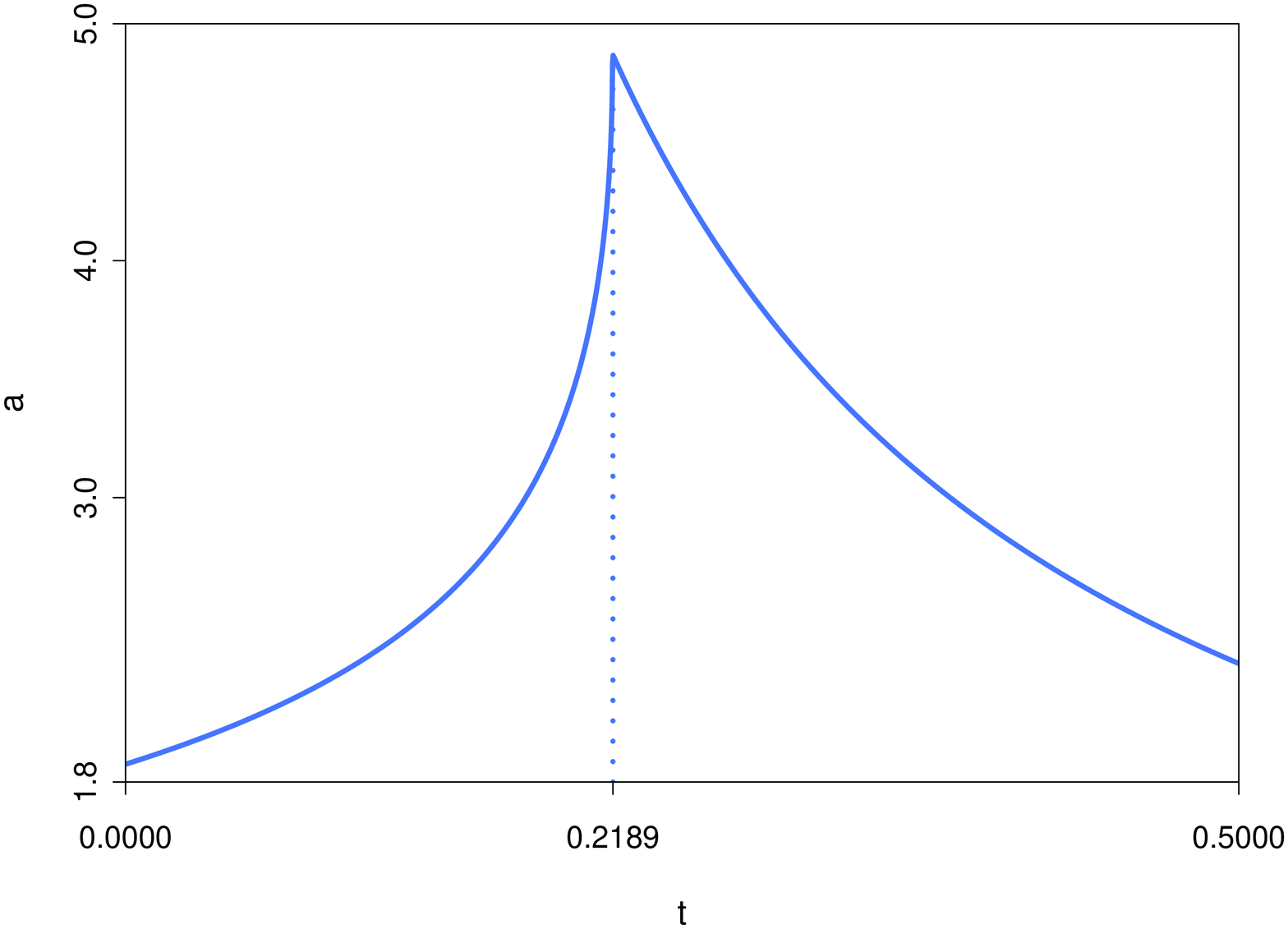}
  \label{fig:roc_MG}
  }
\hspace{8mm}
 \subfloat[Lasota equation]{
 \psfrag{a}[b][b]{\hspace{3mm}Convergence rate ($\text{days}^{-1}$)}
  \includegraphics[width=1.6in,height=2.4in,angle=270]{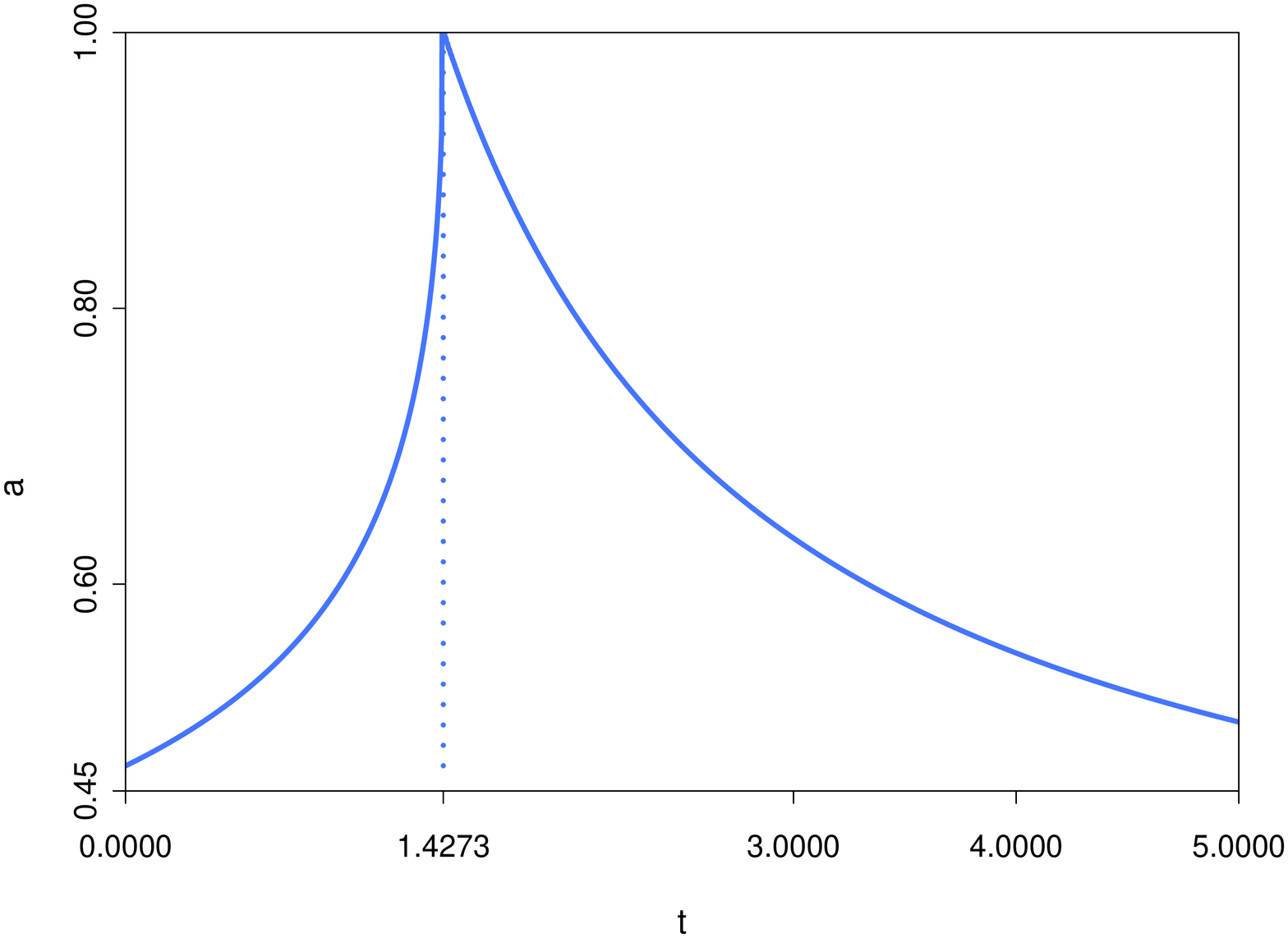}  \label{fig:roc_LS}
  } 
 \end{center}
\caption{Rate of convergence plots for (a) the Mackey-Glass equation~\eqref{eq:mod-m-g}, and (b) the Lasota equation~\eqref{eq:lasota}. For both models, the rate of convergence increases as the feedback delay increases till $\tau^{\ast},$ and decreases for $\tau > \tau^{\ast}.$}
\end{figure*}

We first characterise the dependence of the rate of convergence on the time delay $\tau$ for $a,~b>0$. For this analysis, we consider the linear coefficients $a$ and $b$ of \eqref{eq:general-lin} to be constant. From $p<1$ of \eqref{eq:sc}, we have $-a\tau<1$, \emph{i.e.,} $\tau>-1/a$, a trivial condition for $a>0$, $\tau \geq 0$. From $p < - q$ of \eqref{eq:sc}, we have $ - a < b $ which is also a trivial condition for $a,~b>0$. Using \eqref{eq:sc1}, $p < u_1/\sin (u_1)$ can be written as $\cos (u_1) < 1$, a trivial condition and $ -q < u_1/\sin (u_1)$ can be written as $\cos (u_1) < -a/b$, which is a stricter condition. So, for the case $a,~b>0$, the only relevant stability condition \eqref{eq:sc} is 
\begin{align}
 -q < u_1/\sin (u_1).
\end{align}
As $(- a + \sigma)\tau e^{(a - \sigma)\tau}$ has a maximum $1/e$ at $ (- a + \sigma)\tau~=~1$ and $g(u)$ has a minimum value of $1/e$ at $u=0$, no solution exists for the equation \eqref{eq:msc6} if $b\tau e^ {a\tau} > 1/e$ and for the equation \eqref{eq:msc7} if $b\tau e^ {a\tau} < 1/e.$ For 
$b>0$, $b\tau e^ {a\tau}$ monotonically increases with $\tau$, having a minimum 
\begin{align*}
 (-b/a)(1/e) < 1/e
\end{align*}
for $a,~b>0$ by \eqref{eq:sc}, attained at $ \tau = -1/a$. Thus, there exists $\tau^\ast > -1/a$ such that
\begin{align}
 b\tau^\ast e^ {a\tau^\ast} = \frac{1}{e}.\label{eq:tc}
\end{align}
For $\tau > \tau^*$, no solution exists for equation \eqref{eq:msc6} and for $0 \leq \tau < \tau^\ast$, equation \eqref{eq:msc7} has no solution. For $0 \leq \tau < \tau^*$, let $\sigma_2$ be the solution of equation \eqref{eq:msc6}.
Differentiation of \eqref{eq:msc6} with respect to $\tau$ results in
\begin{align}
\frac{d\sigma}{d\tau} = \frac{b\sigma e^{\sigma \tau}}{1-b\tau e^{\sigma\tau}}.\label{eq:droc}
\end{align}
From \eqref{eq:msc6}, $be^{\sigma \tau}=(-a+\sigma)$. Using this, \eqref{eq:droc} can be re-written as
\begin{align*}
\frac{d\sigma}{d\tau} = \frac{b\sigma e^{\sigma \tau}}{1-(-a+\sigma)\tau}.
\end{align*}
At $ \tau = 0$, $\sigma_1=\infty$ and $\sigma_2 = a+b$. The derivative $ d\sigma_2/d\tau > 0$ if $(- a + \sigma_2)\tau < 1 $, \emph{i.e.} $ \sigma_2 < \sigma_1$. At $ \tau = \tau^\ast$
\begin{align*}
  (- a + \sigma_2)\tau e^{(a - \sigma_2)\tau} =  b\tau e^ {a\tau} = \frac{1}{e}.
\end{align*}
From the above equation, $(- a + \sigma_2)\tau = 1 $, which implies $\sigma_1=\sigma_2$ at $\tau=\tau^*$. For $0 \leq \tau<\tau^*$, $\sigma_2 <\sigma_1$ and $d\sigma_2/d\tau>0$. So, for $0 \leq \tau<\tau^*$, the rate of convergence $\sigma = \min[\sigma_1,\sigma_2] = \sigma_2$, increases as $\tau$ increases.
For $\tau > \tau^*$, from the first condition of \eqref{eq:msc7}, an increase in $\tau$ results in an increase in $u$, thereby resulting in decrease of $\sigma$ as $\tau$ is increased, observable from second condition of \eqref{eq:msc7}. This implies a decrease in the rate of convergence for $\tau>\tau^*$. Since $u_2 > 0$ for $\tau > \tau^\ast$, $u_2/\tan(u_2) < 1$, and $(-a+\sigma_3)\tau < 1$. Thus, $\sigma_3 < \sigma_1$. \\

\noindent \emph{Result}: For $a,~b>0$, the rate of convergence is a monotonically increasing function of $\tau$ given by \eqref{eq:msc6} in the interval $0\leq \tau<\tau^*$, a monotonically decreasing function of $\tau$ given by \eqref{eq:msc7} in the interval $\tau>\tau^*$, $\tau^*$ can be obtained from \eqref{eq:tc}. \\
%
%

\begin{figure*}
 \begin{center}
 \psfrag{7}[b][b]{\scriptsize$7$}
 \psfrag{20}[b][b]{\scriptsize$20$}
 \psfrag{13}[b][b]{\scriptsize$13$}
 \psfrag{0.9}[b][b]{\scriptsize$0.9$}
 \psfrag{7}[b][b]{\scriptsize$7$}
 \psfrag{1}[b][b]{\scriptsize$1$}
 \psfrag{2.4}[b][b]{\scriptsize$2.4$}
 \psfrag{3.8}[b][b]{\scriptsize$3.8$}
 \psfrag{0.5}[b][b]{\scriptsize$0.5$}
 \psfrag{0.9}[b][b]{\scriptsize$0.9$}
 \psfrag{0.1}[b][b]{\scriptsize\hspace{3mm}$0.1$}
 \psfrag{0.230}[b][b]{\scriptsize$0.23$}
 \psfrag{0.240}[b][b]{\scriptsize$0.24$}
 \psfrag{0.252}[b][b]{\scriptsize$0.25$}
 \psfrag{10}[b][b]{\scriptsize\hspace{7mm}$\times \, 10^{-3}$}
  \subfloat[Mackey-Glass equation]{
 \psfrag{a}[b][b]{Feedback delay, $\tau$ (days)}
 \psfrag{t}[b][b]{Non-linearity parameter, $\overline{n}$}
  \psfrag{SSSSSSSSSSSSSS}{\hspace*{-1mm}\scriptsize{Robustly stable}}
  \includegraphics[width=1.6in,height=2.2in,angle=270]{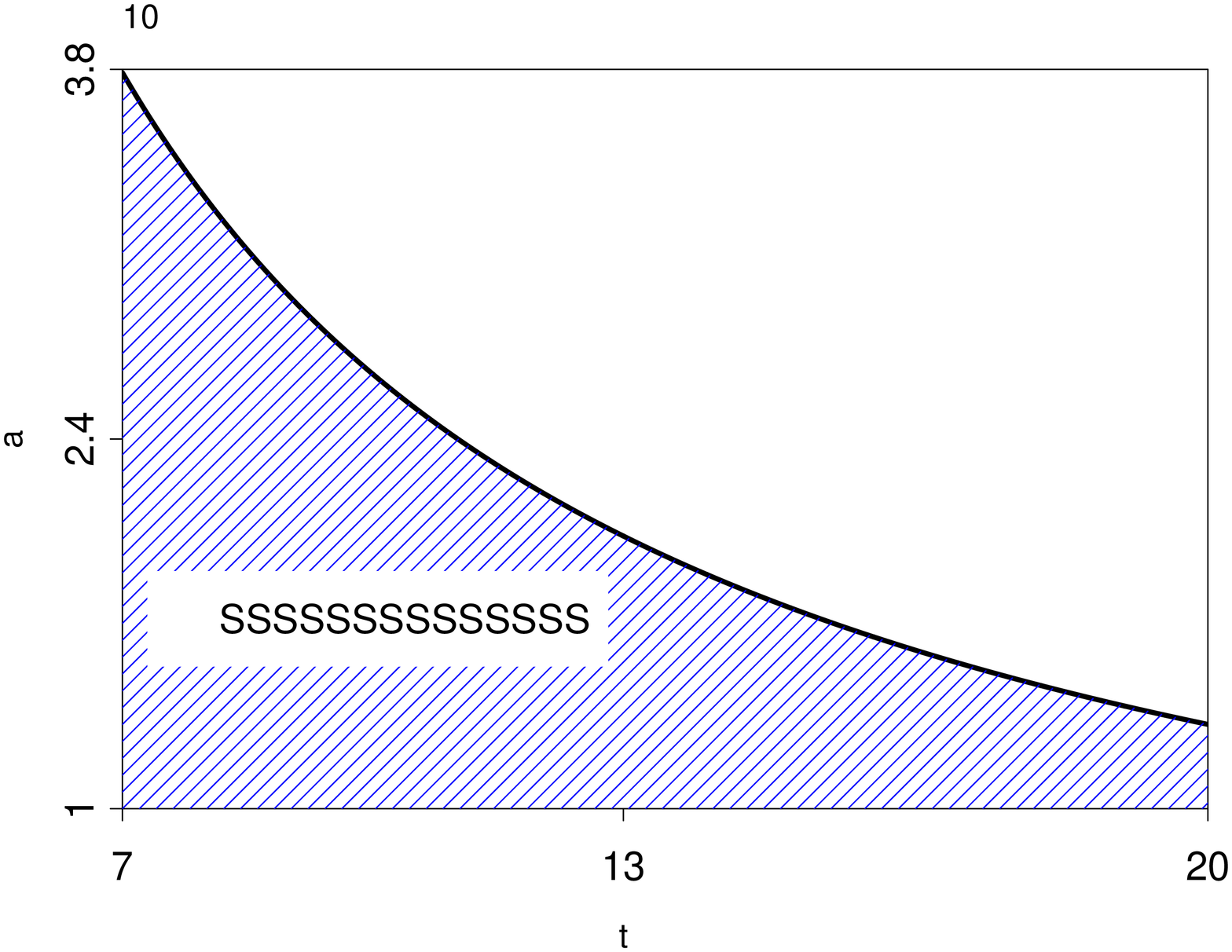}
  \label{fig:robust_MG}
  }
\hspace{8mm}
 \subfloat[Lasota equation]{ 
  \psfrag{SSSSSSSSSSSSSS}{\hspace*{-0.25mm}\scriptsize{Robustly stable}}
  \psfrag{a}[b][b]{Feedback delay, $\tau$ (days)}
   \psfrag{t}[b][b]{Non-linearity parameter, $\underline{n}$}
\includegraphics[width=1.6in,height=2.4in,angle=270]{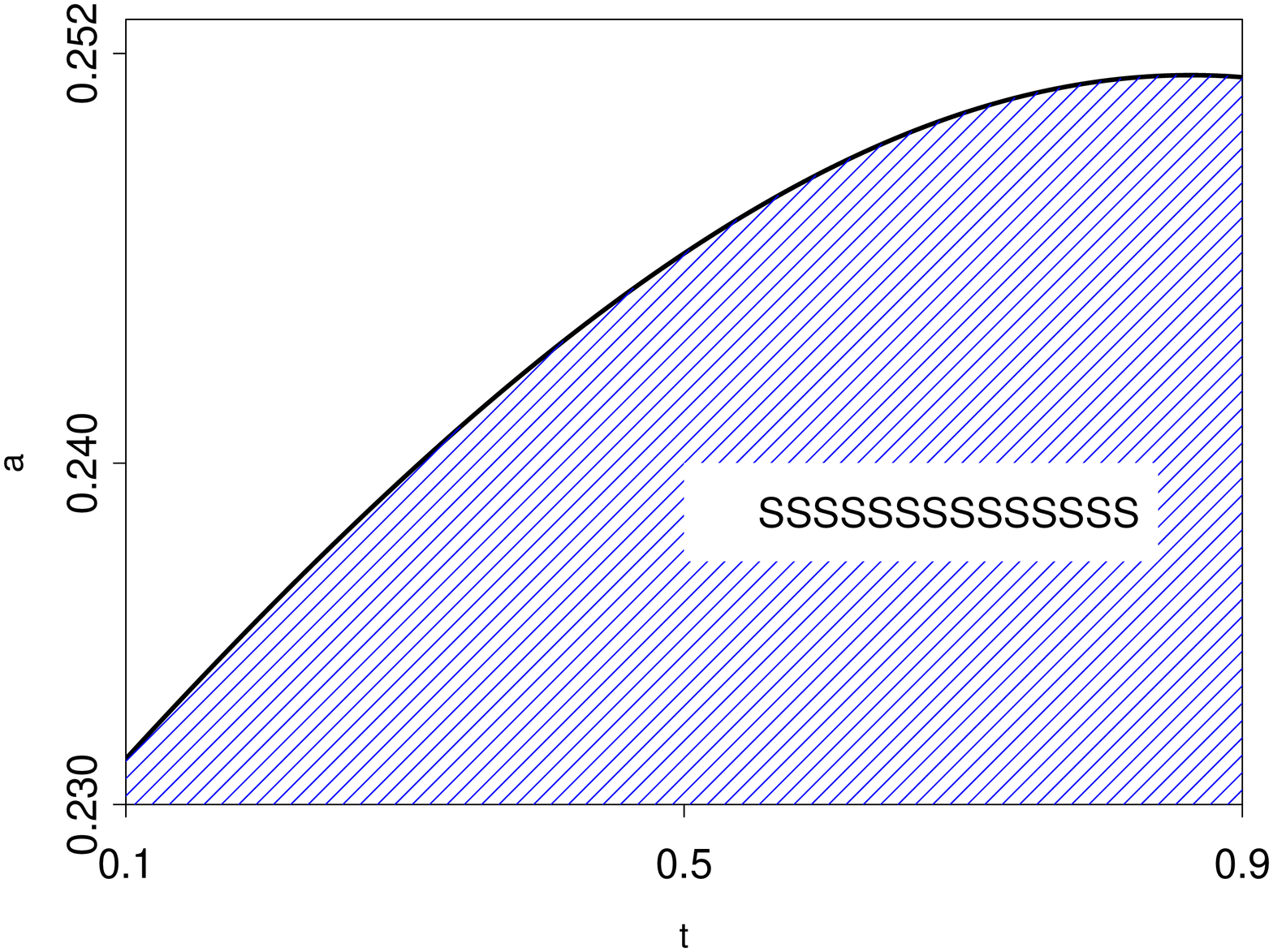}  \label{fig:robust_LS}
  } 
 \end{center}
\caption{Robust stability plots for (a) the Mackey-Glass equation~\eqref{eq:mod-m-g}, and (b) the Lasota equation~\eqref{eq:lasota}. Increasing the worst-case value of the parameter $n$ decreases the value of the delay that stabilises the Mackey-Glass model despite the presence of parametric uncertainties. However, in the Lasota model, we observe the opposite; stabilising the model becomes relatively easier.}
\end{figure*}

\subsection{Mackey-Glass equation}
Using the result stated above, we may characterise the rate of convergence of the solutions for the Mackey-Glass equation as follows. The critical value $\tau^\ast$ solves the equation
\begin{align}
\frac{-\gamma^2}{\beta}\Bigg(1+\bigg(\frac{\beta}{\gamma}-1\bigg)(1-n)\Bigg)\tau^\ast e^{\gamma\tau^\ast} = \frac{1}{e}.
\end{align}
When the feedback delay $\tau < \tau^\ast$, the rate of convergence $\sigma$ solves
\begin{align}
 \frac{-\gamma^2}{\beta}\Bigg(1+\bigg(\frac{\beta}{\gamma}-1\bigg)(1-n)\Bigg) = (\sigma-\gamma)e^{-\sigma\tau},
\end{align}
and when $\tau > \tau^\ast$, $\sigma$ solves
\begin{align}
 \frac{-\gamma^2}{\beta}\Bigg(1+\bigg(\frac{\beta}{\gamma}-1\bigg)(1-n)\Bigg)\tau e^{\gamma\tau} = \frac{u_2}{\sin(u_2)} e^{-u_2/\tan(u_2)},
\end{align}
where $u_2$ can be obtained by solving
\begin{align}
 u_2 = (\sigma - \gamma)\tau \tan(u_2).
\end{align}

We now illustrate this via a numerical example. We use $\beta = 0.8,$ $\gamma = 0.3$ and $n = 10.$ We then obtain $\tau^{\ast} \approx 0.22.$ Thus, we vary the feedback delay from $0$ to $0.5,$ and plot corresponding the rate of convergence of the Mackey-Glass model in Fig.~\ref{fig:roc_MG}. As predicted by the foregoing analysis, the rate of convergence increases till $\tau^{\ast},$ and decreases as it is further increased.

\subsection{Lasota equation}
For the Lasota equation, the critical value of the feedback delay $\tau^\ast$ solves
\begin{align}
 \gamma(x^\ast-n)\tau^\ast e^{\gamma\tau^\ast} = \frac{1}{e},
\end{align}
where the equilibrium $x^\ast$ can be obtained by solving the following transcendental equation
\begin{align*}
 \beta (x^\ast)^{n-1}e^{-x^\ast}=\gamma.
\end{align*}
For the range of feedback delays $\tau < \tau^\ast$, the rate of convergence $\sigma$ is the solution of
\begin{align}
 \gamma(x^\ast-n) = (-\gamma+\sigma) e^{-\sigma\tau}.
\end{align}
When $\tau > \tau^\ast$, $\sigma$ can be obtained by solving the following set of equations
\begin{align}
 \gamma(x^\ast-n)\tau e^{\gamma\tau} =&\, \frac{u_2}{\sin(u_2)} e^{-u_2/\tan(u_2)},\notag\\
  (\sigma - \gamma)\tau \tan(u_2) =&\, u_2.
\end{align}

For the Lasota equation, Fig.~\ref{fig:roc_LS} illustrates the variation in its rate of convergence as a function of the feedback delay. We have used $\beta = 0.4,$ $\gamma = 0.3$ and $n = 0.1,$ which yields $\tau^{\ast} \approx 1.43.$ We notice that the analysis bears out in this numerical example.

\section{Robust stability}
\label{sec:robust_stability}

In practical situations, system parameters are generally not known exactly; they need to be estimated from data. However, the estimation process itself may introduce uncertainties, known as `parametric uncertainties.' In this section, we derive a stability condition (known as the \emph{robust stability condition}) for both models to remain locally stable \emph{despite} paramertic uncertainties.

We begin by assuming that the parameters $\beta,$ $\gamma$ and $n$ in~\eqref{eq:mod-m-g} and~\eqref{eq:lasota} lie in some uncertainty intervals $[\underline{\beta},\overline{\beta}],$ $[\underline{\gamma},\overline{\gamma}]$ and $[\underline{n},\overline{n}]$ respectively. Next, note that~\eqref{eq:general-lin} is of the form~\cite[Equation~$(3)$]{Kharitonov_00}. Hence, a sufficient condition for local stability of both models is~\cite[Lemma~$(3)$]{Kharitonov_00}
\begin{align}
\label{eq:suff_Kha}
b \tau < 1.
\end{align}
Substituting for $b$ from~\eqref{eq:ab}, the above sufficient condition can be written as
\begin{align}
\label{eq:robust_suff1}
-\beta F'(x^\ast) \tau < 1. 
\end{align}
Note the similarity between~\eqref{eq:cond-suff} and~\eqref{eq:robust_suff1}. Using similar manipulations as in Section~\ref{sec:models}, we obtain a sufficient condition for local stability of Mackey-Glass equation as
\begin{align}
\label{eq:MG_new_suff}
\frac{\gamma}{\beta} \Big(n(\beta-\gamma)-\beta\Big) \tau< 1.
\end{align}
Similarly, a sufficient condition for local stability of Lasota equation can be obtained as
\begin{align}
\label{eq:LS_new_suff}
\gamma(x^\ast-n)\tau < 1.
\end{align}
We substitute the ``worst-case'' values of the parameters in~\eqref{eq:MG_new_suff} and~\eqref{eq:LS_new_suff} to obtain a sufficient condition for robust stability of the respective model.

\subsection{Mackey-Glass equation}

A sufficient condition for robust stability the of Mackey-Glass equation is
\begin{align}
\label{eq:robust_MG}
 \frac{\overline{\gamma}}{\underline{\beta}} \Big(\overline{n}(\overline{\beta}-\underline{\gamma})-\underline{\beta}\Big) \tau< 1.
\end{align}
To understand the effect of variation in the parameter $\overline{n},$ we plot~\eqref{eq:robust_MG} as $\overline{n}$ increases from $7$ to $20.$ We choose the uncertainty interval for both $\gamma$ and $\beta$ to be $[0.1,2].$ Fig.~\ref{fig:robust_MG} shows this variation. Note that, as $\overline{n}$ increases, it becomes harder to stabilise the Mackey-Glass model.

\subsection{Lasota equation}
 
A sufficient condition for robust stability of the Lasota equation is
\begin{align}
\label{eq:robust_LS}
\overline{\gamma}(x^\ast-\underline{n})\tau< 1.
\end{align}
We now plot~\eqref{eq:robust_LS} as $\underline{n}$ varies between $0.1$ and $0.9.$ We assume the uncertainty interval for both $\gamma$ and $\beta$ to be $[0.1,2].$ As seen from Fig.~\ref{fig:robust_LS}, in contrast to the Mackey-Glass model, increasing the worst-case value of the parameter $n$ makes it easier to stabilise the model.

Thus, when the feedback delay satisfies~\eqref{eq:robust_MG} or~\eqref{eq:robust_LS}, the corresponding model will remain locally stable, despite the uncertainty in parameter values. This will ensure that, despite the uncertainty introduced by the estimation process, the concentration of blood cells will equilibrate.

\section{Hopf Bifurcation}
\label{sec:Hopf_bifurcation}
In order to study the bifurcation phenomenon in a dynamical system, one needs to identify a suitable bifurcation parameter. 
In both the models, apart from the time delays, there are numerous parameters that could induce the loss of local stability. Therefore, either the time delay or any of the model parameters can be chosen as the bifurcation parameter. However, these model parameters impact the system equilibrium, as observed from the equilibrium conditions. This implies that, variation in any of these parameters would shift the system equilibrium, making it difficult to study the system dynamics and observe the bifurcation phenomenon. Additionally, these parameters may also impact each other. Thus, it may not be ideal to choose either the time delay or any of the model parameters as the bifurcation parameter. To that end, we introduce a non-dimensional parameter $\eta > 0$ which could \emph{act} as the bifurcation parameter. 
This parameter does not influence the system equilibrium, and also enables one to analyse the Hopf bifurcation in both the models in a unified manner. It is assumed that the system has been driven to the Hopf condition, 
where $\eta$ is simply 1. At this point, the parameter $\eta$ acts as the bifurcation parameter by pushing the model
just beyond the edge of stability. So, for both models, $\eta$ can take values between $1$ and $1.05$ which will drive the models into a locally unstable state.  

With the non-dimensional bifurcation parameter, equation \eqref{eq:general} becomes
 \begin{align}
 \dot{x}(t) = \eta\Big(\beta F(x(t-\tau)\big)-\gamma x(t)\Big)\label{eq:general-non-dim}.
\end{align}
Note that the parameter $\eta$ does not effect the equilibrium. We denote the critical value of $\eta$ that satisfies the Hopf condition as $\eta_c$. 
The reader is referred to Appendix B for details of the Hopf bifurcation analysis, and we only present the relevant results 
here.    

\subsection{Mackey-Glass equation} 
To determine the type of the Hopf bifurcation, we have to consider the linear, quadratic and cubic terms in a Taylor series expansion
about the system equilibrium. For the Mackey-Glass equation, these terms are outlined in Table \ref{table:coefficients}.     
Using these terms, and the analysis in Appendix B, we are in a position to determine the type of the Hopf bifurcation 
as $\eta$ crosses its threshold value. We can also comment on the asymptotic orbital stability of the bifurcating limit 
cycles. The expression for the lyapunov coefficient $c_1(0)$ is given in \eqref{eq:c1}, and the expressions for $\mu_2$ and 
$\beta_2$ are given in \eqref{eq:mu} and \eqref{eq:beta}. See Appendix B for the requisite derivations. We now present a numerical example. 

\emph{Numerical Example}: Consider the following parameter values: $\beta=0.8,\gamma=0.3,n=10$. 
For this choice of parameters, using the Hopf condition \eqref{eq:cond-hopf}, 
the system undergoes a Hopf bifurcation when $\tau=\tau_c=1.14$. At this point, $\eta = 1$ and 
as $\eta$ gradually varies beyond 1 the system will find itself in a locally unstable state. 
We set $\eta=1.05$, and computing the required terms in the Hopf analysis, we get the following numerical values: 
\begin{align}
 \mu_2 &= 29.10 > 0,& \beta_2 &=  -35.64 < 0.\notag  
\end{align}
Thus, for this particular choice of parameters, the Hopf bifurcation is \emph{super-critical} and the emergent limit cycles are asymptotically orbitally stable. 
\begin{table*}
\centering
\caption{Coefficients in the Taylor series expansion of the Mackey-Glass and Lasota equations.}
\begin{tabular}{l l l}\hline
\\ [-2.5mm]
 Coefficient & Mackey-Glass equation& Lasota equation\\\hline\\[-1ex]
 $\xi_x$ & $\displaystyle-\gamma$ & $\displaystyle-\gamma$\\[2ex]
 $\xi_y$ & $\displaystyle\frac{\gamma^2}{\beta}\bigg((1-n)\frac{\beta}{\gamma}+n\bigg)$ & $\displaystyle\gamma(n-x^\ast)$\\[2ex]
 $\xi_{yy}$ &$\gamma n(\beta-\gamma)\Big((\beta-\gamma)(n-1)-\gamma(n+1)\Big)\bigg/2\beta^2 x^\ast $& $\displaystyle \gamma\Big((x^\ast)^2-2x^\ast n + n(n-1)\Big)\bigg/2x^\ast$\\[2ex]
 $\xi_{yyy}$&$\gamma n(\beta-\gamma)\Big((\beta-\gamma)^2(1-n^2)$&$\displaystyle\gamma\Big((n-x^\ast)^3+n\big(2-3(n-x^\ast)\big)\Big)\Big/6(x^\ast)^2$\\[2ex]
 &$+\,\gamma(\beta-\gamma)(4n^2+2)+\gamma^2(1-n^2)\Big)\Big/6\beta^3(x^\ast)^2 $&\\[2ex]\hline\\
 Equilibrium Condition &$(x^\ast)^n = \frac{\beta}{\gamma}-1$ & $\beta(x^\ast)^{n-1}e^{-x^\ast}=\gamma$\\[2ex]
 \hline\label{table:coefficients}
\end{tabular}
\end{table*}

\begin{figure}
 \centering
 \psfrag{x}[b][b]{Bifurcation parameter, $\eta$}
 \psfrag{y}[b][b]{$x(t)$ (cells/kg)}
 \psfrag{0.9}[b][b]{\scriptsize$0.9$}
 \psfrag{1.0}[b][b]{\scriptsize$1.0$}
 \psfrag{1.1}[b][b]{\scriptsize$1.1$}
 \psfrag{1.2}[b][b]{\scriptsize$1.2$}
 \includegraphics[width=1.6in,height=2.4in,angle=270]{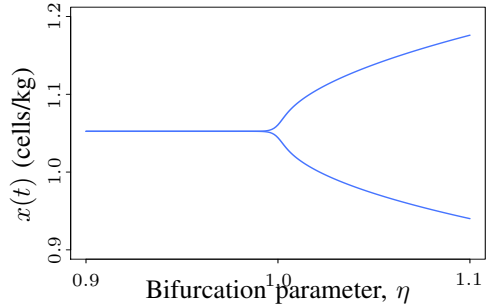}
 \caption{Bifurcation diagram for the Mackey-Glass equation \eqref{eq:mod-m-g}.  It can be seen that, as $\eta$ varies, the system undergoes a change from a stable equilibrium to limit cycles. The amplitude of the bifurcating periodic solutions is observed to increase as the bifurcation parameter $\eta$ increases. The parameter values used were: $\beta=0.8,\gamma=0.3,n=10$.}
 \label{fig:bifrcation_diagram_MG}
\end{figure}

\begin{figure*}
 \begin{center}
 \psfrag{xt}{$x(t)$}
  \psfrag{xT}{$x(t-\tau)$}
   \psfrag{e}{$\scriptsize\text{Equilibrium point}$}
   \psfrag{0.7}[b][b]{\scriptsize$0.7$}
   \psfrag{0.9}[b][b]{\scriptsize$0.9$}
   \psfrag{1.1}[b][b]{\scriptsize$1.1$}
   \psfrag{1.3}[b][b]{\scriptsize$1.3$}
   \psfrag{1.0}[b][b]{\scriptsize$1.0$}
   \psfrag{1.2}[b][b]{\scriptsize$1.2$}
 \subfloat[Stable equilibrium, $\tau < \tau_c$]{
  \includegraphics[width=1.6in,height=2.2in,angle=270]{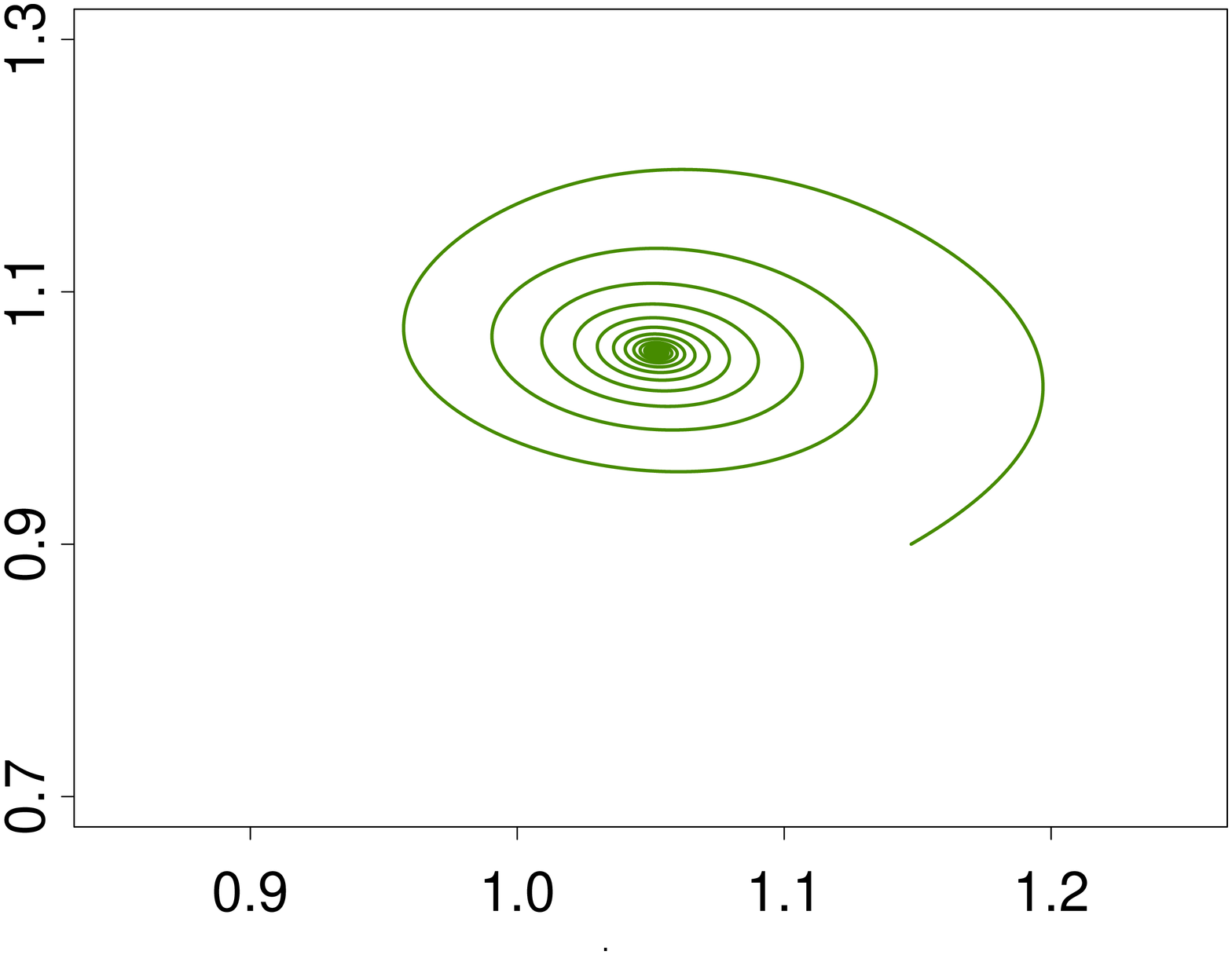}
  \label{fig:convergence_MG}
  } 
  \hspace{8mm}
  \subfloat[Stable limit cycle, $\tau > \tau_c$]{
  \includegraphics[width=1.6in,height=2.2in,angle=270]{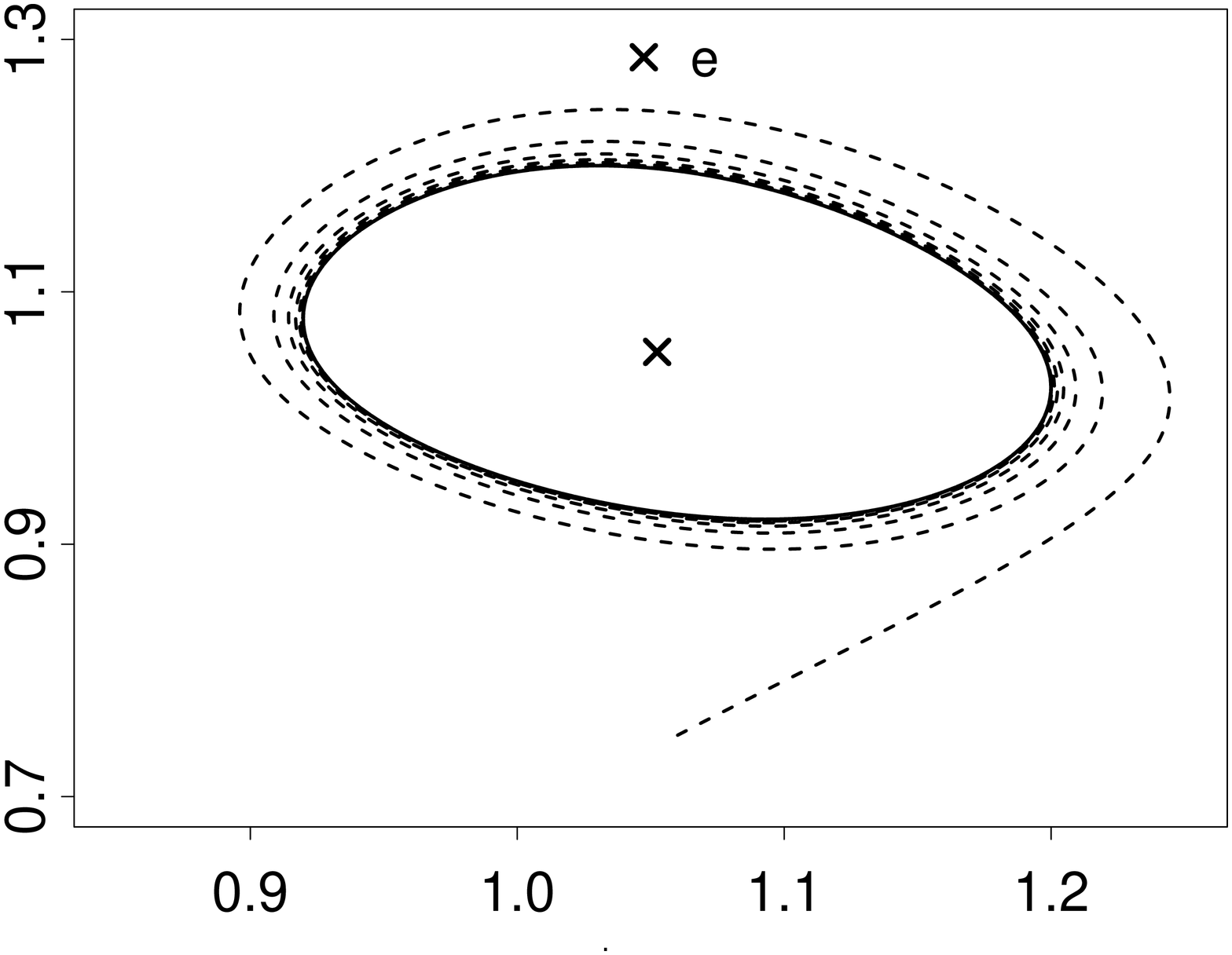}
  \label{fig:lim_cycle_MG}
  }
 \end{center}
\caption{Phase portraits of the Mackey-Glass equation \eqref{eq:mod-m-g} showing (a) convergence of trajectories to the equilibrium, $x^\ast = 1.0524$, for $\tau=1<\tau_c$, and (b) the emergence of a limit cycle for $\tau=1.3>\tau_c$. We can see that the system undergoes a qualitative change from a \emph{stable equilibrium} to a \emph{limit cycle}. The parameter values used were: $\beta=0.8,\gamma=0.3,n=10$.}
\end{figure*}

A bifurcation diagram for the Mackey-Glass equation is shown in Figure \ref{fig:bifrcation_diagram_MG}. For the construction of this diagram, 
we used the same parameter values as outlined above. The Hopf bifurcation occurs at $\eta = 1$, and as this parameter varies
we witness the emergence of limit cycles, whose amplitudes are plotted in the bifurcation diagram. For $\eta=1.05$, we explicitly established that the Hopf bifurcation is super-critical. To show the transition from a stable equilibrium to a limit cycle, we plot the phase portraits of the Mackey-Glass equation with variations in the time delay. In Figure \ref{fig:convergence_MG} we see convergence to the equilibrium point, and in Figure \ref{fig:lim_cycle_MG} we can clearly observe the presence of a limit cycle. Variations in other parameters of the system, including the parameter $\eta$, could also have been used to construct the phase plots. 
\subsection{Lasota equation}
As we have presented the Hopf analysis, with the parameter $\eta$, in a very general manner we are in a position to 
conduct a similar exercise with the Lasota equation. The associated Taylor series expansions are tabulated in Table \ref{table:coefficients}. We now present a numerical example.

\begin{figure}
 \centering
  \psfrag{x}[b][b]{Bifurcation parameter, $\eta$}
 \psfrag{y}[b][b]{$x(t)$ (cells/kg)}
 \psfrag{0.9}[b][b]{\scriptsize$0.9$}
 \psfrag{1.0}[b][b]{\scriptsize$1.0$}
 \psfrag{1.1}[b][b]{\scriptsize$1.1$}
 \psfrag{1.5}[b][b]{\scriptsize$1.5$}
 \psfrag{2.0}[b][b]{\scriptsize$2.0$}
 \psfrag{2.5}[b][b]{\scriptsize$2.5$}
 \psfrag{3.0}[b][b]{\scriptsize$3.0$}
 \includegraphics[width=1.6in,height=2.4in,angle=270]{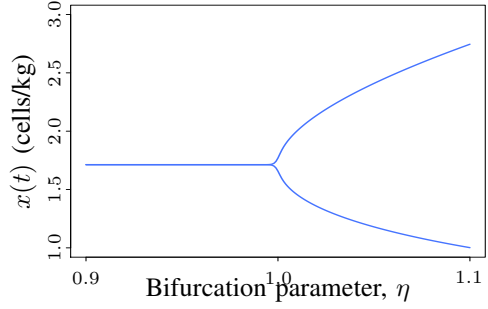}
 \caption{Bifurcation diagram for the Lasota equation \eqref{eq:lasota}.  It can be seen that, as $\eta$ varies, the system undergoes a change from a stable equilibrium to limit cycles. The amplitude of the bifurcating periodic solutions is as shown above. The parameter values used were: $\beta=0.9,\gamma=0.1,n=0.1$.}
 \label{fig:bifrcation_diagram_LS}
\end{figure}

\begin{figure*}
 \begin{center}
 \psfrag{xt}{$x(t)$}
  \psfrag{xT}{$x(t-\tau)$}
   \psfrag{e}{$\scriptsize\text{Equilibrium point}$}
   \psfrag{0}[b][b]{\scriptsize$0$}
   \psfrag{1}[b][b]{\scriptsize$1$}
   \psfrag{2}[b][b]{\scriptsize$2$}
   \psfrag{3}[b][b]{\scriptsize$3$}
   \psfrag{4}[b][b]{\scriptsize$4$}
 \subfloat[Stable equilibrium, $\tau < \tau_c$]{
  \includegraphics[width=1.6in,height=2.2in,angle=270]{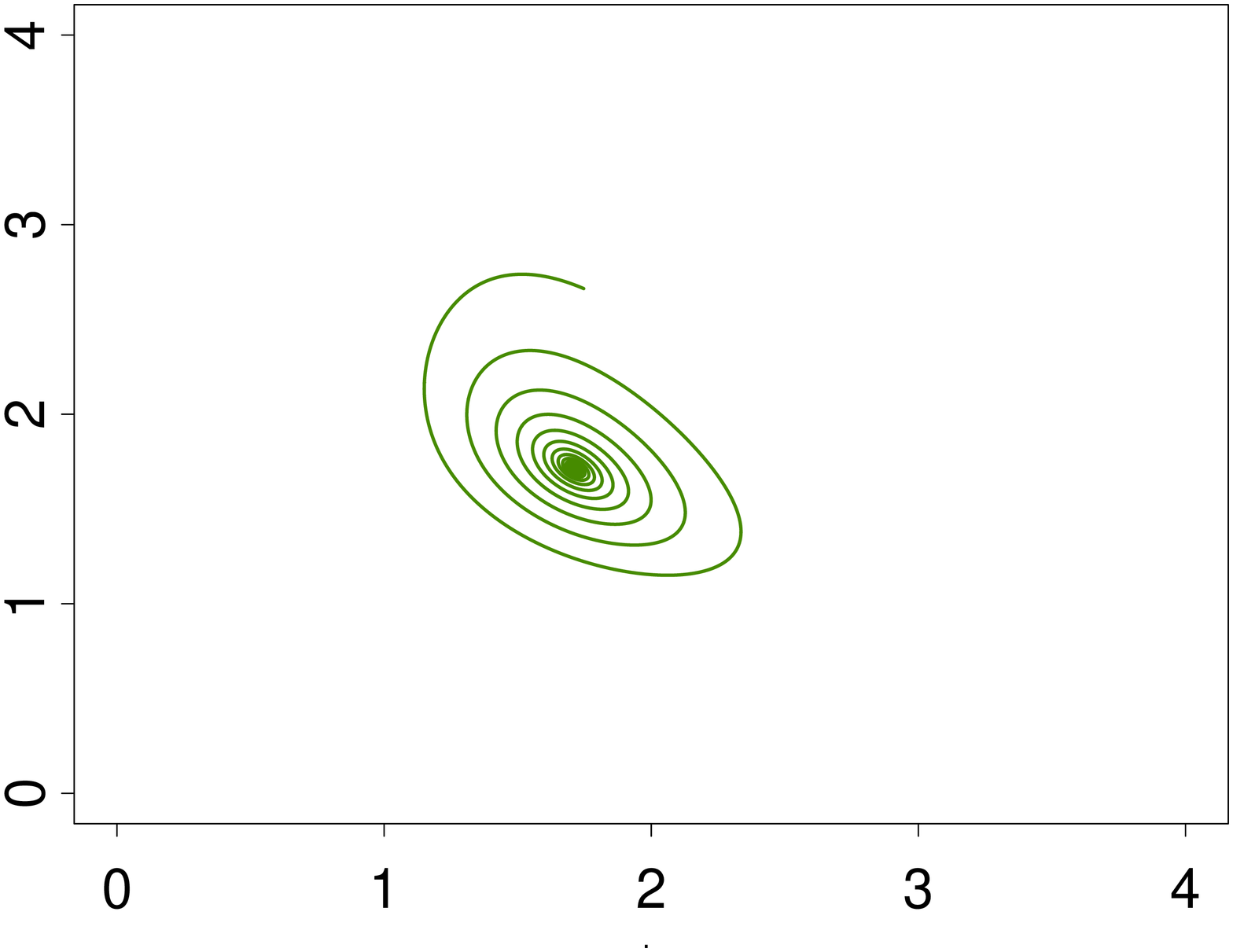}
  \label{fig:convergence_LS}
  } 
  \hspace{8mm}
  \subfloat[Stable limit cycle, $\tau > \tau_c$]{
  \includegraphics[width=1.6in,height=2.2in,angle=270]{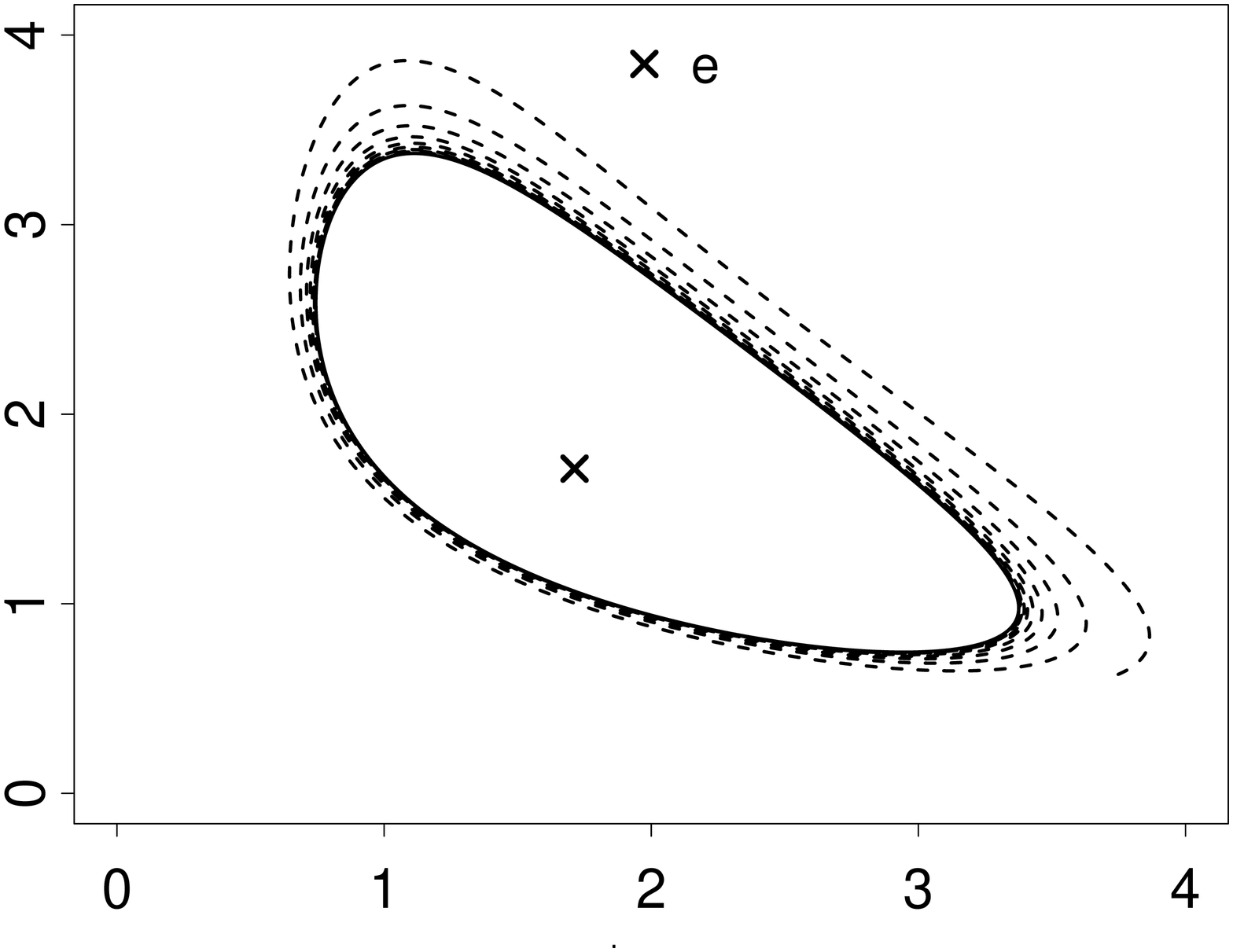}
  \label{fig:lim_cycle_LS}
  }
 \end{center}
\caption{Phase portraits of the Lasota equation \eqref{eq:lasota} showing (a) convergence of trajectories to the equilibrium, $x^\ast = 1.71$, for $\tau=13.69<\tau_c$, and (b) the emergence of a limit cycle for $\tau=21.69>\tau_c$. We can see that the system undergoes a qualitative change from a \emph{stable equilibrium} to a \emph{limit cycle}. The parameter values used were: $\beta=0.9,\gamma=0.1,n=0.1$.}
\end{figure*}

\emph{Numerical Example:} For the following parameter values $\beta=0.9,\gamma=0.1,n=0.1$, the critical time delay at which the Hopf bifurcation occurs is $\tau=\tau_c=  17.69$. Again, at the Hopf condition $\eta = 1$, and we set $\eta = 1.05$. Using the 
analysis in the Appendix, we get the following numerical values:
\begin{align}
 \mu_2 &=   0.8072 > 0,& \beta_2 &=  -0.0398 < 0.\notag  
\end{align}
Thus, for this choice of parameters, we can conclude that the Hopf bifurcation is \emph{super-critical}, and the limit cycles are asymptotically orbitally stable. The bifurcation diagram (Fig.~\ref{fig:bifrcation_diagram_LS}) and phase portraits (Fig.~\ref{fig:convergence_LS}-\ref{fig:lim_cycle_LS}) of the Lasota equation indicate the emergence of limit cycles in the blood cell concentration when system stability is lost. The numerical simulations were conducted using the scientific computing software MATLAB.

\section{Contributions}
\label{sec:contributions}
We studied two mathematical models for human physiological processes: (i) the Mackey-Glass equation which models haematopoiesis, and (ii) the Lasota equation which models erythropoiesis. These models are known to capture both normal and pathological behaviour in the respective processes~\cite{Glass_79}. In our work, we used control- and bifurcation- theoretic tools as applied to non-linear time-delayed systems in order to understand the system dynamics and also the qualitative behaviour of the underlying physiological processes.

First, we analysed the local stability of the two models. Stability implies that the blood cell concentration would converge to the equilibrium, which signifies normal behaviour. Here, we derived a simple sufficient condition for local stability, that can guide design of system parameters such that system stability is ensured. We then derived the necessary and sufficient condition for local stability. This condition gives strict bounds on system parameters for stable operation. We also outlined conditions for non-oscillatory convergence of the solutions, which could ensure design of parameters such that blood cell concentration equilibrates without exhibiting oscillations. We discussed the trade-offs between the feedback delay and (i) the dependence parameter $\beta$, and (ii) non-linearity parameter $n$ for local stability and non-oscillatory convergence of solutions of both models.

While guaranteeing system stability ensures that the blood cell concentration reaches the equilibrium, it does not yield insight into the rate of convergence. Therefore, we studied the rate of convergence of the solutions for both models. We outlined closed-form expressions for the rate of convergence as a function of the feedback delay. The rate is seen to be an increasing function of the delay up to a critical value of the delay, and then monotonically decreases as the delay increases beyond the said critical value.

As physiological parameters cannot be estimated precisely, it is desirable to obtain conditions for system stability in the presence of parameter uncertainty. Therefore, we derived conditions for both models to be robustly stable in the presence of parametric uncertainties. We illustrated the dependence of the feedback delay required to ensure robustly stable operation on the non-linearity parameter $n$ via some stability plots.  

Further, we explicitly showed that when the necessary and sufficient condition for local stability is violated, the system transits into instability via a Hopf bifurcation. This signifies the emergence of un-damped limit cycle oscillations in the blood cell concentration, which indicates pathological behaviour in the human body. It would be desirable to understand if the emergent limit cycles are orbitally stable/unstable. To that end, we conduct a detailed Hopf bifurcation analysis. Using the center manifold theory and Poincar\'{e} normal forms, we outlined an analytical framework to determine the type of the Hopf bifurcation and asymptotic orbital stability of the limit cycles. The use of the outlined framework is illustrated through numerical examples. Our study yields insights into the variation in blood cell concentration under normal and pathological conditions. This may, in turn, guide therapy for dynamical diseases.

\subsection{Avenues for further research}
\label{sec:future_work}
Of the two models that we studied, the Mackey-Glass equation is reasonably well studied in the literature, while the Lasota equation needs to be understood better. In our work, we only considered local properties of the two models, \emph{i.e}, the system behaviour in a small neighbourhood of the equilibrium. Global stability and global bifurcation properties of the two models, specially those of the Lasota equation, are yet to be explored. Also, one could explore the impact of heterogeneous time delays\textemdash that could capture the variation in time taken for the release of mature blood cells into the blood stream across different parts of the body\textemdash on the system dynamics in terms of stability, rate of convergence and bifurcation.
\appendix

\section{Local Stability Analysis}
In Appendix A, we present the local stability analysis of a general non-linear delay differential equation. Consider the following equation:
\begin{align}
 \dot{x}(t) = \eta f\big(x(t),x(t-\tau)\big),\label{eq:gen_mod}
\end{align}
where $f$ has an equilibrium point denoted by $(x^\ast,y^\ast)$ and $\eta> 0,\tau \geq 0$. We define $u(t) = x(t) - x^\ast$, and linearise equation \eqref{eq:gen_mod} to obtain
\begin{align}
 \dot{u}(t) = -\eta a u(t) - \eta b u(t-\tau),\label{eq:lin_gen_mod}
\end{align}
where $a,b>0$ are given by $a = -f_x|_{(x^\ast,y^\ast)}$ and $b = -f_y|_{(x^\ast,y^\ast)}$ and $b > a$.
Looking for exponential solutions, we have the following characteristic equation
\begin{align}
 \lambda + \eta a + \eta b e^{-\lambda\tau} = 0.\label{eq:gen_char}
\end{align}
We now derive a sufficient condition for local stability of system~\eqref{eq:gen_mod} using the Nyquist stability criteria. 
From the characteristic equation~\eqref{eq:gen_char}, one may derive the loop transfer function of the linearised system~\eqref{eq:lin_gen_mod} as
\begin{align*}
 L(\lambda) = \frac{\eta\, b e^{-\lambda\tau}}{\lambda + \eta a},
\end{align*}
which, at $\lambda = i\omega$, becomes
\begin{align*}
 L(i\omega) = \frac{\eta\, b e^{-i\omega\tau} }{-i\omega\tau +\eta a}.
\end{align*}
According to the Nyquist stability criteria, the linearised system~\eqref{eq:lin_gen_mod} would be stable if the magnitude of the loop transfer function is upper bounded by unity, at the point of crossover, \emph{i.e.} $$|L(\theta)| < 1, \,\,\,\text{when}\,\,\,\, \angle L(\theta) = \pi.$$
We first seek the cross-over frequency, \emph{i.e.}, the frequency for which phase of the loop transfer function is $\pi$. This yields
\begin{align}
 \tan(\omega\tau) = \frac{-\omega}{\eta\, a}.\label{eq:gen_phase_cond}
\end{align}
Using the bound on the magnitude of the loop transfer function, we get 
\begin{align*}
 |L(i\omega)| = \frac{\eta\, b}{\sqrt{\omega^2 + \eta^2 a^2}} < 1,
\end{align*}
which can be written as 
\begin{align}
 \frac{\eta\, b\tau}{\omega\tau}\sin(\omega\tau) < 1. \label{eq:gen_magnitude_cond}
\end{align}
We may derive the required condition for local stability by applying the above bound on the supremum of the function of the left-hand side of inequality~\eqref{eq:gen_magnitude_cond}. From equations\eqref{eq:gen_phase_cond} and~\eqref{eq:gen_magnitude_cond}, we gather that 
$\tan(\omega\tau) < 0$ and $\sin(\omega\tau) > 0$. This implies that $\pi/2 < \omega\tau < \pi$. Therefore, the function on the left-hand side of inequality~\eqref{eq:gen_magnitude_cond} attains its supremum when $\omega\tau = \pi/2$, consequently $\sin(\omega\tau)=1$. This yields the following sufficient condition for local stability of system~\eqref{eq:gen_mod}
\begin{align}
 \eta b\tau < \frac{\pi}{2}.\label{eq:gen_suff}
\end{align}

Following the analysis outlined in \cite{Raina_05}, we derive the necessary and sufficient condition for local stability of system~\eqref{eq:gen_mod} as
\begin{align}
 \eta\tau \sqrt{b^2-a^2} < \cos^{-1}\Big(\frac{-a}{b}\Big).\label{eq:gen_necc}
\end{align}
It has been shown that the model transits into instability via a Hopf bifurcation, see \cite{Raina_05}. However, in such models a Hopf bifurcation could occur when either the non-dimensional parameter $\eta$, or the delay $\tau$ or any of the system parameters is varied. 
Hence, any of these can be used as the bifurcation parameter. Recall that, we wish to choose $\eta$ as the bifurcation parameter as it enables us to capture the effect of variation in any of the system parameters.
It has been shown that the system undergoes a Hopf bifurcation at
\begin{align}
 \eta \tau \sqrt{b^2 - a^2} = \cos^{-1}\Big(\frac{-a}{b}\Big),\label{eq:gen_hopf}
\end{align}
with a period of $2\pi\tau/\cos^{-1}(-a/b)$.
We denote the value of $\eta$ and $\tau$ satisfying equation \eqref{eq:gen_hopf} as $\eta_c$ and $\tau_c$ respectively. Having established that the system undergoes a Hopf bifurcation, we now present the analysis that enables us to address questions about the type of Hopf bifurcation and the stability of emergent limit cycles.
 
\section{Hopf Bifurcation Analysis}
Recalling that $u(t)=x(t)-x^\ast$, a Taylor series expansion of equation \eqref{eq:gen_mod} including the linear, quadratic and cubic terms is 
\begin{align}
 \dot{u}(t) =&\,\, \eta\xi_x u(t) + \eta\xi_y u(t-\tau)+ \eta\xi_{yy}u^2(t-\tau) + \eta\xi_{yyy}u^3(t-\tau)+\cdots,\label{eq:gen_expand}
\end{align}
where the coefficients of the linear, quadratic and cubic terms are 
\begin{align}
 \xi_{x} &= f_{x}|_{(x^\ast,y^\ast)},&\xi_{y}&=f_{y}|_{(x^\ast,y^\ast)},\notag\\
  \xi_{yy}&=\frac{1}{2}f_{yy}|_{(x^\ast,y^\ast)},& \xi_{yyy}&=\frac{1}{6}f_{yyy}|_{(x^\ast,y^\ast)}.\notag
\end{align}
Note that we retain only those terms that appear in the Taylor series expansion of the Mackey-Glass and Lasota equations. 
We now perform the requisite analysis to determine the type of the Hopf bifurcation. We closely follow the style of analysis outlined in \cite{Raina_05,Hassard_81}.
Consider the following autonomous delay differential equation
\begin{align}
 \dot{u}(t) = \mathcal{L_\mu}u_t + \mathcal{F}(u_t,\mu),\label{eq:stand_form}
\end{align}
where for $t>0,\mu\in\mathbb{R}$ and $\tau>0$, 
\begin{align}
 u_{t}(\theta) = u(t+\theta), \hspace{4mm} u \mathrm{:}\hspace{1pt}[-\tau,0]\rightarrow \mathbb{R}, \hspace{4mm}\theta \in [-\tau,0].\notag
\end{align}
We define $\mathcal{L_\mu}$, a one-parameter family of linear operators, as $\mathcal{L_\mu}\mathrm{:}\hspace{2pt}C[-\tau,0]\rightarrow\mathbb{R}$. The operator $\mathcal{F}(u_t,\mu):\hspace{2pt}C[-\tau,0]\rightarrow\mathbb{R}$ contains the non-linear terms.
Assume that $\mathcal{F}(u_{t},\mu)$ is analytic and that $\mathcal{F}$ and ${\mathcal{L}_{\mu}}$ depend on the \emph{bifurcation parameter} $\eta=\eta_c+\mu$ for small $|\mu|$. Note that \eqref{eq:gen_expand} is a type of the form of \eqref{eq:stand_form}. We now cast \eqref{eq:stand_form} into the following form
\begin{align}
 \frac{d}{dt} u_{t} = \mathcal{A}(\mu)u_{t} + \mathcal{R}u_{t}\label{eq:operator_form}
\end{align}
as equation \eqref{eq:operator_form} contains $u_{t}$ rather than both $u$ and $u_{t}$. We first transform the linear terms $(d/dt)u(t)=\mathcal{L}_{\mu}u_{t}$. We use the Riesz representation theorem that guarantees the existence of an $n \times n$ matrix-valued function $\rho(.,\mu):\hspace{2pt}[-\tau,0]\rightarrow\mathbb{R}^{n^{2}}$, such that each component of $\rho$ has bounded variation and for all $\phi \in  C[-\tau,0]$ 
\begin{align*}
 \mathcal{L}_{\mu}\phi = \int_{-\tau}^{0}\,\mathrm{d}\rho (\theta,\mu)\phi(\theta).
\end{align*}
In particular
\begin{align}
 \mathcal{L}_{\mu}u_{t} = \int_{-\tau}^{0}\,\mathrm{d}\rho (\theta,\mu) u(t+\theta),\label{eq:lin_op}
\end{align}
where
\begin{align*}
  \mathrm{d}\rho(\theta,\mu) = \,\,\eta\big(\xi_{x}\delta(\theta)+ \xi_{y}\delta(\theta+\tau)\big)\mathrm{d}\theta,
\end{align*}
and $\delta(\theta)$ is the Dirac-delta function, that satisfies equation \eqref{eq:lin_op}.
We now define, for $\phi\in C^{1}[-\tau,0]$, the following operators
\begin{align}
  \mathcal{A}(\mu)\phi(\theta) &= \begin{cases}\begin{array}{ll}\frac{\mathrm{d}\phi(\theta)}{\mathrm{d}\theta},& \theta \in [-\tau,0)\\ \int_{-\tau}^{0}\,\mathrm{d}\rho(s,\mu)\phi(s)\equiv \mathcal{L}_{\mu}\phi,&\theta=0,\end{array}\end{cases}\label{eq:Aoperator}\\
   \mathcal{R}\phi(\theta)&= \begin{cases} \begin{array}{ll} 0,& \theta \in [-\tau,0)\\ \mathcal{F}(\phi,\mu),&\theta = 0.\end{array}\end{cases}\label{Roperator}
  \end{align}
Then, as $\mathrm{d}u_t/\mathrm{d}\theta \equiv \mathrm{d}u_t/\mathrm{d}t$, equation \eqref{eq:stand_form} becomes \eqref{eq:operator_form}.
 We now proceed to determine the coefficients required for the Hopf bifurcation analysis.
 Note that $\eta=\eta_c+\mu$ is the bifurcation parameter under consideration. As $\eta=\eta_c$ at the point of bifurcation, we set $\mu = 0 $ in order to compute the required terms at the point of bifurcation.
Let $q(\theta)$ be the eigenfunction for $\mathcal{A}(0)$ corresponding to $\lambda(0)$, namely
\begin{align*}
 \mathcal{A}(0)q(\theta) = i\omega_{0}q(\theta),
\end{align*}
and define the \emph{adjoint} operator $\mathcal{A}^{\ast}(0)$ as
\begin{align*}
 \mathcal{A}^{\ast}(0)\alpha(s) = \begin{cases} \begin{array}{ll}-\frac{\mathrm{d}\alpha(s)}{\mathrm{d}s},&s\in (0,\tau]\\\int_{-\tau}^{0}\mathrm{d}\rho^{\tau}(t,0)\alpha(-t)&s=0\end{array}\end{cases}
\end{align*}
where $\rho^{T}$ denotes the transpose of $\rho$. The domain of $\mathcal{A}^\ast$ being $C^1[0,\tau]$.
If $q(\theta)$ is the eigenvector of $\mathcal{A}$ corresponding to the eigenvalue $\lambda(0)$, then $\bar{\lambda}(0)$ is an eigenvalue of $\mathcal{A}^\ast$, and 
\begin{align*}
 \mathcal{A}^{\ast}q^{\ast} = -i\omega_{0}q^{\ast},
\end{align*}
 where $q^{\ast}$ is some nonzero vector. For $\phi  \in  C[-\tau,0]$ and $\psi  \in  C[0,\tau]$, define an inner product
\begin{align}
 \langle\psi,\phi\rangle =&\,\, \bar{\psi}(0)\cdot\phi(0) - \int_{\theta=-\tau}^{0} \int_{\zeta=0}^{\theta}\bar{\psi}^{T}(\zeta-\theta)\,d\eta(\theta)\phi(\zeta)d\zeta.\label{eq:inner_prod}
\end{align}
Let $q(\theta)=e^{i\omega_0\theta}$ and $q^{\ast}(s) = De^{i\omega_0s}$ be the eigenvectors of $\mathcal{A}$ and $\mathcal{A}^\ast$ corresponding to the eigenvalues $i\omega_0$ and $-i\omega_0$. We then find $D$, such that $\langle q^\ast,q\rangle = 1$ and $\langle q^\ast,\bar{q}\rangle = 0$, as 
\begin{align}
 D = \frac{1}{1+\tau\eta\xi_y e^{i\omega_0\tau}}.
\end{align}
For $u_{t}$, a solution of \eqref{eq:operator_form} at $\mu = 0$, define
\begin{align*}
  z(t) &= \langle q^{\ast},u_{t}\rangle ,\hspace{3mm} \text{and} \\
 \mathsf{w}(t,\theta) &= u_{t}(\theta)- 2 \text{Re}\big(z(t)q(\theta)\big).
\end{align*}
Then, on the manifold, $C_{0},\mathsf{w}(t,\theta) = \mathsf{w}\big(z(t),\bar{z}(t),\theta\big)$ where
\begin{align}
 \mathsf{w}(z,\bar{z},\theta) = \mathsf{w}_{20}(\theta)\frac{z^{2}}{2}+ \mathsf{w}_{11}(\theta)z\bar{z} + \mathsf{w}_{02}(\theta)\frac{\bar{z}^{2}}{2}+\cdots.\label{eq:defn_w}
\end{align}
In the directions of the eigenvectors $q^{\ast}$ and $\bar{q}^{\ast}$,  $z$ and $\bar{z}$ are local coordinates for $C_{0}$ in $C$ respectively. Equation \eqref{eq:stand_form} can now be reduced to an ordinary differential equation for a single complex variable on $C_{0}$. At $\mu = 0$, we have
\begin{align}
   z'(t) &= \langle q^{\ast},\mathcal{A}u_{t} + \mathcal{R}u_{t}\rangle\notag\\
   &= i\omega_{0}z(t) + \bar{q}^{\ast}(0)\cdot \mathcal{F}\Big(\mathsf{w}(z,\bar{z},\theta)+ 2\text{Re}\big(z(t)q(\theta)\big)\Big)\notag\\
   &= i\omega_{0}z(t) + \bar{q}^{\ast}(0)\cdot \mathcal{F}_{0}(z,\bar{z})\notag\\
   &=i\omega_{0}z(t) + g(z,\bar{z}),\label{eq:zprime}
\end{align}
where $g(z,\bar{z})$ can be expanded in powers of $z$ and $\bar{z}$ as 
\begin{align}
 g(z,\bar{z}) &= \bar{q}^{\ast}(0)\cdot\mathcal{F}_{0}(z,\bar{z})\notag\\
   &= g_{20}\frac{z^{2}}{2} + g_{11}z \bar{z} + g_{02}\frac{\bar{z}^{2}}{2}+ g_{21}\frac{z^{2}\bar{z}}{2}+ \cdots.\label{eq:defn_g}
\end{align}
We now need to determine the $\mathsf{w}_{ij}(\theta)$ in equation \eqref{eq:defn_w}. Following \cite{Hassard_81} we write
\begin{align*}
 \mathsf{w}' = u'_{t} - z'q - \bar{z}'\bar{q},
\end{align*}
and using \eqref{eq:operator_form} and \eqref{eq:zprime} we obtain
\begin{equation*}
 \mathsf{w}' = \begin{cases}
               \begin{array}{l l}
                \mathcal{A}\mathsf{w} - 2\text{Re}\big(\bar{q}^\ast(0)\cdot\mathcal{F}_0q(\theta)\big),& \theta \in [-\tau,0)\\
                \mathcal{A}\mathsf{w} - 2\text{Re}\big(\bar{q}^\ast(0)\cdot\mathcal{F}_0q(0)\big)+\mathcal{F}_0, & \theta = 0
               \end{array}
              \end{cases}
\end{equation*}
which can be written as 
\begin{equation}
 \mathsf{w}' = \mathcal{A}\mathsf{w} + H(z,\bar{z},\theta),\label{eq:wprime1}
\end{equation}
using \eqref{eq:zprime}, where
\begin{equation}
  H(z,\bar{z},\theta) = H_{20}(\theta)\frac{z^{2}}{2} + H_{11}(\theta)z\bar{z} + H_{02}(\theta)\frac{\bar{z}^{2}}{2} + \cdots.\label{eq:defnH}
\end{equation}
Now, on the manifold $C_{0}$, near the origin
 \begin{equation}
  \mathsf{w}' = \mathsf{w}_{z}z' + \mathsf{w}_{\bar{z}}\bar{z}'.\label{eq:wprime}
 \end{equation} 
 Use equations \eqref{eq:defn_w} and \eqref{eq:zprime} to replace $\mathsf{w},z'$ (and their conjugates by their power series expansion) and equating this with \eqref{eq:wprime1}, we get 
 \begin{align}
  (2i\omega_{0}- \mathcal{A})\mathsf{w}_{20}(\theta) &= H_{20}(\theta),\label{eq:H20}\\
 -\mathcal{A}\mathsf{w}_{11}(\theta) &= H_{11}(\theta),\label{eq:H11}\\
 -(2i\omega_{0}+ \mathcal{A})\mathsf{w}_{02}(\theta) &= H_{02}(\theta)\label{eq:H02},
 \end{align}
 as in \cite{Hassard_81}. We now expand the non-linear terms in equation \eqref{eq:gen_expand} using the following 
 \begin{align}
 u_{t}(\theta) =&\,\, \mathsf{w}(z,\bar{z},\theta)+ zq(\theta) + \bar{z}\bar{q}(\theta)\notag\\
 =&\,\, \mathsf{w}_{20}(\theta)\frac{z^{2}}{2} + \mathsf{w}_{11}(\theta)z\bar{z} + \mathsf{w}_{02}(\theta)\frac{\bar{z}^{2}}{2} + ze^{i\omega_{0}\theta} + \bar{z}e^{-i\omega_{0}\theta}+ \cdots, \label{eq:defn_u}
 \end{align}
 to obtain $u_{t}(0)$ and $u_{t}(-\tau)$. Retaining only the coefficients of $z^2,z\bar{z},\bar{z}^2,z^2\bar{z}$, we have
\begin{align}
  u_{t}^{2}(-\tau)=&\,\, z^{2}e^{-2i\omega_{0}\tau}+ \bar{z}^{2}e^{2i\omega_{0}\tau} + 2z\bar{z} + z^{2}\bar{z}\big(2e^{-i\omega_{0}\tau}\mathsf{w}_{11}(-\tau)+ e^{i\omega_{0}\tau}\mathsf{w}_{20}(-\tau)\big)+ \cdots,\notag\\
  u_{t}^{3}(-\tau) =&\,\, 3z^{2}\bar{z}e^{-i\omega_{0}\tau}+\cdots.\label{eq:non-linear_terms}
\end{align}
Using equation \eqref{eq:defn_g} and \eqref{eq:non-linear_terms}, we get
\begin{align}
   g_{20} =&\hspace{2pt} 2\bar{q}^{\ast}(0)\eta\xi_{yy} e^{-2i\omega_{0}\tau},\notag\\
   g_{11} =&\hspace{2pt} 2\bar{q}^{\ast}(0)\eta\xi_{yy},\notag\\
   g_{02} =&\hspace{2pt} 2\bar{q}^{\ast}(0)\eta\xi_{yy}e^{2i\omega_{0}\tau},\notag\\
   g_{21} =&\hspace{2pt} \bar{q}^{\ast}(0)\eta\bigg(\xi_{yy}\Big(4\mathsf{w}_{11}(-\tau)e^{-i\omega_{0}\tau}+ 2\mathsf{w}_{20}(-\tau)e^{i\omega_{0}\tau}\Big) + 6\xi_{yyy}e^{-i\omega_{0}\tau}\bigg).\label{eq:gcoeff}
  \end{align}
In the expression for $g_{21}$ we have $\mathsf{w}_{11}(0),\mathsf{w}_{11}(-\tau),\mathsf{w}_{20}(0)$ and $\mathsf{w}_{20}(-\tau)$ which need to be evaluated. Now, for $\theta \in [-\tau,0) $
  \begin{align*}
   H(z,\bar{z},\theta) =& -2\text{Re}\big(\bar{q}^{\ast}(0)\cdot \mathcal{F}_{0} q(\theta)\big)\\
   =& -2 \text{Re} \big(g(z,\bar{z})q(\theta)\big)\\
   =& -\left(g_{20}\frac{z^{2}}{2} + g_{11}z\bar{z} + g_{02}\frac{\bar{z}^{2}}{2}+\cdots \right)q(\theta) - \left(\bar{g}_{20}\frac{\bar{z}^{2}}{2} + \bar{g}_{11}z\bar{z} + \bar{g}_{02}\frac{z^{2}}{2}+\cdots \right)\bar{q}(\theta).
  \end{align*}
From \eqref{eq:defnH}, we have
  \begin{align*}
  H_{20}(\theta) &= -g_{20}q(\theta)-\bar{g}_{02}\bar{q}(\theta),\\
  H_{11}(\theta) &= -g_{11}q(\theta)-\bar{g}_{11}\bar{q}(\theta).\\
  \end{align*}
Using equations \eqref{eq:Aoperator}, \eqref{eq:H20} and \eqref{eq:H11}, we get
\begin{align}
   \mathsf{w}'_{20}(\theta) &= 2i\omega_{0}\mathsf{w}_{20}(\theta) + g_{20}q(\theta)+ \bar{g}_{02}\bar{q}(\theta),\label{eq:w20prime}\\
   \mathsf{w}'_{11}(\theta) &= g_{11}q(\theta) + \bar{g}_{11}\bar{q}(\theta).\label{eq:w11prime}
  \end{align}
  Solving equations \eqref{eq:w20prime} and \eqref{eq:w11prime}, we obtain 
  \begin{align}
   \mathsf{w}_{20}(\theta) &= -\frac{g_{20}}{i\omega_{0}}q(0)e^{i\omega_{0}\theta}-\frac{\bar{g}_{02}}{3i\omega_{0}}\bar{q}(0)e^{-i\omega_{0}\theta} + E_{1}e^{2i\omega_{0}\theta},\label{eq:w20}\\
   \mathsf{w}_{11}(\theta) &= \frac{g_{11}}{i\omega_{0}}q(0)e^{i\omega_{0}\theta} - \frac{\bar{g}_{11}}{i\omega_{0}}\bar{q}(0)e^{-i\omega_{0}\theta} + E_{2},\label{eq:w11}
  \end{align}
  where $E_{1},E_{2}$ need to be determined. For $H(z,\bar{z},0) = -2\text{Re}\big(\bar{q}^{\ast}\cdot \mathcal{F}_{0} q(0)\big) + \mathcal{F}_{0}$, we have 
  \begin{align}
   H_{20}(0) =& -g_{20}q(0)-\bar{g}_{02}\bar{q}(0)+2\eta\xi_{yy}e^{-2i\omega_{0}\tau},\label{eq:H20z}\\
   H_{11}(0) =& -g_{11}q(0)-\bar{g}_{11}\bar{q}(0)+2\eta\xi_{yy}.\label{eq:H11z}
  \end{align}
  From equations \eqref{eq:Aoperator}, \eqref{eq:H20} and \eqref{eq:H11}, we get
  \begin{align}
  \eta\xi_{x}\mathsf{w}_{20}(0) + \eta\xi_{y}\mathsf{w}_{20}(-\tau)-2i\omega_{0}\mathsf{w}_{20}(0) =&\, \hspace{1pt}g_{20}q(0)+\bar{g}_{02}\bar{q}(0)-2\eta\xi_{yy}e^{-2i\omega_{0}\tau},\label{eq:sol1}\\
   \eta\xi_{x}\mathsf{w}_{11}(0) + \eta\xi_{y}\mathsf{w}_{11}(-\tau) =&\, g_{11}q(0)+\bar{g}_{11}\bar{q}(0)- 2\eta\xi_{yy}.\label{eq:sol2}
  \end{align}
  We find $\mathsf{w}_{20}(0),\mathsf{w}_{20}(-\tau),\mathsf{w}_{11}(0)$ and $\mathsf{w}_{11}(-\tau)$ using equations \eqref{eq:w20} and \eqref{eq:w11} and substitute in equations \eqref{eq:sol1} and \eqref{eq:sol2} to evaluate $E_{1},E_{2}$. We finally obtain
  \begin{align*}
   E_{1} &= \frac{\Theta_{1}}{\eta\xi_{x}+\eta\xi_{y}e^{-2i\omega_{0}\tau}-2i\omega_{0}},\\
   E_{2} &= \frac{\Theta_{2}}{\eta(\xi_{x}+\xi_{y})},
  \end{align*}
  where
  \begin{align*}
   \Theta_{1} =& \hspace{1pt}(\eta\xi_{x}-2i\omega_{0})\bigg(\frac{g_{20}}{i\omega_{0}} + \frac{\bar{g}_{02}}{3i\omega_{0}}\bigg) + \eta\xi_{y}\bigg(\frac{g_{20}}{i\omega_{0}}e^{-i\omega_{0}\tau} + \frac{\bar{g}_{02}}{3i\omega_{0}}e^{i\omega_{0}\tau}\bigg)+ g_{20}q(0)+\bar{g}_{02}\bar{q}(0)\\ &-2\eta\xi_{yy}e^{-2i\omega_{0}\tau},\\
   \Theta_{2} =& -\eta\xi_{x}\bigg(\frac{g_{11}}{i\omega_{0}}-\frac{\bar{g}_{11}}{i\omega_{0}}\bigg)-\eta\xi_{y}\bigg(\frac{g_{11}}{i\omega_{0}}e^{-i\omega_{0}\tau} - \frac{\bar{g}_{11}}{i\omega_{0}}e^{i\omega_{0}\tau}\bigg)+ g_{11}q(0)+\bar{g}_{11}\bar{q}(0)- 2\eta\xi_{yy}.
  \end{align*}
  All quantities required for the computations associated for the stability analysis of the Hopf bifurcation have been derived. We now use the following expressions to analyse the type of the Hopf bifurcation \cite{Hassard_81}
  \begin{align}
   c_{1}(0) &= \frac{i}{2\omega_{0}}\left(g_{20}g_{11} - 2|g_{11}|^{2} - \frac{1}{3}|g_{02}|^{2}\right) + \frac{g_{21}}{2},\label{eq:c1}\\
   \mu_{2} &= \frac{-\text{Re}\big(c_{1}(0)\big)}{\alpha'(0)},\label{eq:mu}\\
   \beta_{2} &= 2\text{Re}\big(c_{1}(0)\big)\label{eq:beta},
  \end{align}
  where $g_{20},g_{11},g_{02},g_{21}$ are defined by equation \eqref{eq:gcoeff}, and $c_{1}(0)$ is the lyapunov coefficient. The term $\alpha'(0)=\text{Re}(d\lambda/d\eta)$ evaluated at $\eta=\eta_c$. The period of the bifurcating solutions is $2\pi/\omega_0$.
  We now state the conditions that enable us to characterise the type of Hopf bifurcation, and determine the stability of the bifurcating periodic solutions.
  \begin{enumerate}[i.]
   \item The sign of $\mu_2$ determines the type of Hopf bifurcation. The Hopf bifurcation is \emph{super-critical} if $\mu_2>0$ and \emph{sub-critical} if $\mu_2<0$.
  \item The sign of $\beta_2$, which is the Floquet exponent, determines the \emph{asymptotic orbital stability} of the bifurcating periodic solutions. The periodic solutions are stable if $\beta_2 < 0$ and unstable if $\beta_2 > 0$.
  \end{enumerate}






\end{document}